\documentclass[10pt]{iopart}

\usepackage[utf8]{inputenc} %
\usepackage[T1]{fontenc}    %
\usepackage{hyperref}       %
\usepackage{url}            %
\usepackage{booktabs}       %
\usepackage{amsfonts}       %
\usepackage{nicefrac}       %
\usepackage{microtype}      %
\usepackage[table, x11names]{xcolor}         %
\usepackage[most]{tcolorbox} %
\definecolor{bgcolor}{rgb}{0.97, 0.97, 0.97}

\usepackage{siunitx} %
\usepackage{fontawesome5}  %

\usepackage{amsmath} %
\usepackage{mathtools}

\usepackage{url} %
\usepackage{float}

\usepackage{caption}
\usepackage{subcaption}

\usepackage{listings} %

\usepackage{booktabs}

\definecolor{amaranth}{rgb}{0.9, 0.17, 0.31}
\colorlet{green}{green!20}
\colorlet{yellow}{yellow!60}
\colorlet{red}{red!30}

\usepackage{setspace}
\usepackage[linesnumbered,ruled,vlined]{algorithm2e}

\SetKwComment{Comment}{\color{green!80!black!190}// }{}

\SetKwProg{Function}{function}{}{}
\SetKw{Continue}{continue}

\DontPrintSemicolon
\SetAlFnt{\small}
\SetAlgorithmName{Pseudocode}{algorithmautorefname}

\usepackage{tikz}
\usetikzlibrary{decorations.pathreplacing,calc}

\usepackage{enumitem}

\usepackage{siunitx} %

\usepackage{diagbox} %

\usepackage{multirow}  

\usepackage{longtable} %

\usepackage{array}
\newcolumntype{?}[1]{!{\vrule width #1}}
\def\showcomments{} %

\ifx\showcomments\undefined
\newcommand{\CV}[1]{}
\newcommand{\YP}[1]{}
\else
\newcommand{\CV}[1]{\textcolor{red}{[CV: #1]}}
\newcommand{\YP}[1]{\textcolor{purple}{[YP: #1]}}
\fi

\usepackage[absolute,overlay]{textpos}
\usepackage{times}

\usepackage{scalerel}
\newcommand\vfrac[2]{\ThisStyle{%
  \setbox0=\hbox{$\SavedStyle#1#2$}%
  \setbox2=\hbox{$\SavedStyle X$}%
  \ifdim\ht0>\ht2\setlength{\ht0}{\ht2}\fi%
  #1\mathord{\stretchto{\raisebox{2.3\LMpt}{$\SavedStyle/$}}{\ht0}}#2}}

\usepackage{cite}
\usepackage{placeins}

\begin{document}

\DeclarePairedDelimiterX{\infdivx}[2]{(}{)}{%
  #1\;\delimsize\|\;#2%
}

\ioptwocolmod
\twocolumn[{
\begin{@twocolumnfalse}
\title[Plasma State Monitoring and Disruption Characterization using Multimodal VAEs]{Plasma State Monitoring and Disruption Characterization using Multimodal VAEs}
\vspace{-2pt}
\author{Yoeri Poels$^{1,2}$, Alessandro Pau$^1$, Christian Donner$^3$, Giulio Romanelli$^3$, Olivier Sauter$^1$, Cristina Venturini$^1$, Vlado Menkovski$^2$, the TCV team$^4$ and the WPTE team$^5$}
\address{$^1$École Polytechnique Fédérale de Lausanne (EPFL), Swiss Plasma Center (SPC), CH-1015 Lausanne, Switzerland}
\address{$^2$Eindhoven University of Technology (TU/e), Mathematics and Computer Science, NL-5600MB Eindhoven, The Netherlands}
\address{$^3$Swiss Data Science Center (SDSC), ETH Zürich \& EPFL, CH-8092 Zürich \& CH-1015 Lausanne, Switzerland}
\address{$^4$See author list of B. P. Duval \textit{et al.} 2024 \textit{Nucl. Fusion} \textbf{64} 112023}
\address{$^5$See author list of E. Joffrin \textit{et al.} 2024 \textit{Nucl. Fusion} \textbf{64} 112019}
\vspace{-2.6pt}
\ead{yoeri.poels@epfl.ch}
\vspace{1.4pt}
\begin{indented}
\item[]April 2025
\end{indented}
\textbf{Abstract}\\
When a plasma disrupts in a tokamak, significant heat and electromagnetic loads are deposited onto the surrounding device components. These forces scale with plasma current and magnetic field strength, making disruptions one of the key challenges for future devices. Unfortunately, disruptions are not fully understood, with many different underlying causes that are difficult to anticipate. Data-driven models have shown success in predicting them, but they only provide limited interpretability. On the other hand, large-scale statistical analyses have been a great asset to understanding disruptive patterns. In this paper, we leverage data-driven methods to find an interpretable representation of the plasma state for disruption characterization. Specifically, we use a latent variable model to represent diagnostic measurements as a low-dimensional, latent representation. We build upon the Variational Autoencoder (VAE) framework, and extend it for (1) continuous projections of plasma trajectories; (2) a multimodal structure to separate operating regimes; and (3) separation with respect to disruptive regimes. Subsequently, we can identify continuous indicators for the disruption rate and the disruptivity based on statistical properties of measurement data. The proposed method is demonstrated using a dataset of approximately 1600 TCV discharges, selecting for flat-top disruptions or regular terminations. We evaluate the method with respect to (1) the identified disruption risk and its correlation with other plasma properties; (2) the ability to distinguish different types of disruptions; and (3) downstream analyses. For the latter, we conduct a demonstrative study on identifying parameters connected to disruptions using counterfactual-like analysis. Overall, the method can adequately identify distinct operating regimes characterized by varying proximity to disruptions in an interpretable manner.

\vspace{.13cm}
\hrule
\vspace{.13cm}
\end{@twocolumnfalse}
}\vspace{-2.9cm}]

\renewcommand*{\thefootnote}{\arabic{footnote}}

\maketitle

\section{Introduction}\label{sec:intro}
A disruption in a tokamak corresponds to the rapid loss of plasma control. The plasma is no longer confined, terminating the discharge and rapidly depositing the plasma's thermal and electromagnetic energy onto the surrounding vessel. This phenomenon can cause severe machine damages~\cite{Jepu_2019,Jepu_2024}, especially in future devices~\cite{lehnen2015,strait2019}, making disruptions one of the key challenges for the exploitation of tokamaks in future power plants~\cite{fasoli2023fusion}. Unfortunately, the exact physical processes involved in disruptions are not fully understood, many different trajectories can lead to disruptive conditions, and it is not fully determined how to best predict their onset~\cite{wesson2003,disruptionpredictionITPA,Hender2007,deVries2009,deVries2011}. 

Past experiments have generated vast datasets capturing both disruptive and non-disruptive plasma behavior. Large-scale statistical analyses have been a cornerstone for the understanding of disruptions~\cite{deVries2009,deVries2011,Gerasimov2020}. Simultaneously, expressive statistical methods, such as deep neural networks (NNs)~\cite{LeCun2015}, have shown much success in the task of predicting disruptions~\cite{rea2019,zheng2023disr,zhu021disr,katesharbeck2019,Vega2022,Aymerich2022,aledda2015}. However, these disruption prediction methods often only provide limited interpretability. Instead, we aim to leverage similar expressive statistical methods with an explicit focus on finding interpretable representations of the plasma operational space, rather than directly mapping observables to a prediction in a black-box manner.

We propose a method for plasma state monitoring using a \textit{latent variable model}. Here, the goal is to find an abstract, low-dimensional representation that models a large quantity of discharges, each defined by a time series of signal data. In other words, we aim to automatically identify an abstract phase space of a tokamak's operational space. Specifically, we leverage machine learning methods to learn the transformation from data space to this so-called latent space, which is simultaneously optimized to accurately represent the original measurements while providing separability w.r.t.\ disruptive regimes. We propose the method as a complement to prediction methods, instead focusing on enhancing analysis and understanding.

We leverage the framework of Variational Autoencoders (VAEs)~\cite{Kingma2014,rezende2014}. A VAE models a data distribution $p(x)$ under the assumption that it is generated by a latent variable $z$, and models the relationship between data space and latent space using conditional distributions parametrized by neural networks. We adapt this framework with three properties in mind: (1) a continuous projection of sequential measurements---for tracking discharges as they evolve; (2) a multimodal latent variable---for separating the operational space into discrete regimes; all while (3) providing separability w.r.t.\ disruptive regimes. We implement (1) using a dynamic formulation for the encoder distribution~\cite{Girin2021}, modeling the time derivative of the latent variable. For (2), we utilize a multimodal, Gaussian mixture-based prior distribution for $z$~\cite{dilokthanakul2017deep,tomczak2022}. To improve training dynamics, we propose extensions to the model architecture and its optimization. Finally, (3) is achieved by learning a `disruption risk' variable as a function of $z$. 

Due to the smooth structure of the learned manifold and its separability w.r.t.\ disruptive dynamics, we can use it to identify continuous analogues of the disruption rate and the disruptivity~\cite{deVries2009,disruptionpredictionITPA} (as detailed in Section~\ref{sec:problem}). Additionally, in this initial study, we constrain $z$ to be 2-dimensional, allowing for easier interpretability of the manifold and its relation to physical parameters. The proposed method is demonstrated with a dataset of approximately 1600 TCV discharges. %
We characterize flat-top dynamics, consequently selecting for flat-top disruptions or regular terminations, and additionally filter on discharges reaching a diverted X-point configuration. TCV provides a challenging testbed given its large variability in plasma scenarios~\cite{duval2024} and its sensitivity to fast disruptions connected to MHD limits~\cite{defuseiaea,marchioni2024,GTurri_2008}.

We validate the identified disruption risk variable both quantitatively and qualitatively. Specifically, we evaluate its calibration w.r.t.\ the disruption rate, compare it to the disruptivity, and identify its correspondence to known operational limits. In this context, we evaluate the ability to automatically distinguish clusters of distinct types of disruptions, as found in ITER Baseline Scenario (IBL) experiments~\cite{labit2024}, density limits~\cite{sieglin2025} and negative triangularity scenarios~\cite{coda2022}. Additionally, we correlate the identified states in latent variable $z$ with plasma properties such as the confinement state~\cite{poels2025ldh} and disruption precursors~\cite{defuseiaea}. Lastly, we conduct a proof-of-principle study of using the latent variable model to automatically identify disruption-related parameters, by identifying counterfactual pairs of (non-)disrupting discharges and the times-of-interest therein.

Adjacent latent variable-based methods have utilized Generative Topographic Mapping~\cite{Bishop1998,pau2019,Aymerich2021} and Self-Organizing Maps (SOM)~\cite{kohonen1982,Raffaele2012,Aymerich2024} to characterize disruptive regions in the tokamak operational space. In contrast, we utilize the VAE for its expressivity in learning the latent variable utilizing neural networks, alongside its flexibility in shaping the optimization target. Closest to our setting are~\cite{Wei2021} and~\cite{buli2025jet}, both also building upon the VAE framework. We differentiate through an explicit focus on learning a multimodal, state-based representation, allowing for better clustering of distinct operating regimes. Notably, there are also ongoing efforts towards automated detection of chains-of-events connected to disruptions~\cite{Sabbagh2023decaf,defuseiaea}. We consider the proposed method complementary to such efforts, providing an extra tool to this end. In short, our contributions can be summarized as follows:
\begin{itemize}
    \item We develop a method to automatically identify a low-dimensional, multimodal, interpretable representation of the operational space of TCV, optimized to identify operational limits w.r.t.\ disruptions.
    \begin{itemize}
        \item We build upon existing VAE-based approaches, and propose extensions to the model architecture and its optimization to encourage multimodality in the learned posterior distribution.
        \item The method automatically clusters distinct regimes, and can be used to identify continuous analogues for the disruption rate and disruptivity of these parameter spaces.
    \end{itemize}
    \item We build a database of approximately 1600 TCV discharges containing either (a) flat-top disruptions or (b) regular terminations, covering a representative sample of the TCV operational space. 
    \item We extensively evaluate the proposed method both quantitatively and qualitatively. We evaluate (1) latent space properties and their connection to disruption metrics and events; (2) the clustering of distinct operating regimes; and (3) using the model for downstream analyses. For the latter, we conduct a proof-of-principle study of identifying disruption-related parameters using counterfactual-like analysis.
\end{itemize}

\section{Problem Formulation}\label{sec:problem}
Disruptions in tokamaks are known to be caused by a wide array of phenomena~\cite{deVries2009,deVries2011,Maraschek2018}. Various operational limits have been identified, such as the Greenwald limit for the plasma density~\cite{Greenwald2002} or the low-$q$ current limit (Kruskal-Shafranov limit), relating to the ratio of toroidal to poloidal magnetic field components~\cite{kruskal1958,shafranov1958,mhdbook}. Many different plasma conditions are associated with heightened risk of instabilities eventually leading to a disruption~\cite{wesson2003,disruptionpredictionITPA}. Statistical analyses of past disruptions can aid the understanding of disruptive boundaries and disruption causes~\cite{deVries2011}, and simultaneously present the opportunity for learning a plasma state representation for integration into advanced control schemes~\cite{vu2021}. Consequently, the automatic identification of patterns present in large datasets of disrupting discharges can aid efforts towards better understanding of disruptions and contribute to achieving more robust device operations. We aim to learn a nonlinear low-dimensional representation of high-dimensional observations, efficiently compressing down the vast amount of experimental data to its core patterns.

We define the problem setting as modeling a dataset of tokamak discharges $\mathbf{x} \in X \subseteq \mathbb{R}^{U \times T^{in}}$, for $U$ input signals and $T^{\textit{in}}$ time samples. These discharges are modeled using a latent variable $\mathbf{z} \in \mathbb{R}^{d \times T^z}$ of $d$ dimensions, for latent trajectories of $T^z$ timesteps. A prior distribution is assumed for $p(\mathbf{z})$, allowing us to model the data distribution $p(\mathbf{x})$ as $p(\mathbf{x}, \mathbf{z}) = p(\mathbf{x}|\mathbf{z})p(\mathbf{z})$. Since we model the data as timesteps, we aim to map realizations of latent variable $\mathbf{z}$ to signals at a single timestep $t_m$, i.e., we want to model:
\begin{align}
    p(\mathbf{x}^{t_m}|\mathbf{z}^{t_m}),
\end{align}
for a \textit{decoder} distribution $p$ mapping from latent space to data space\footnote{We omit explicit indexing on the signals for $\mathbf{x}$, i.e. $\mathbf{x}^\mathbf{u}$, for conciseness, since we always operate on all input signals.}. However, we do not want to treat individual timesteps as independent, but rather model the plasma state as a trajectory in the latent space. Consequently, we aim to learn a distribution where each latent state depends on the previous state (Markov assumption):
\begin{align}\label{eq:encproblemstatement}
    q(\mathbf{z}^{t_1}, \ldots, \mathbf{z}^{t_m}|\mathbf{z}^{t_0},\mathbf{x}^{\mathbf{t}\leq{t_m}}) = \prod_{i=1}^{m}q(\mathbf{z}^{t_i}|\mathbf{z}^{t_{i-1}}, \mathbf{x}^{\leq t_i}),
\end{align}
for an \textit{encoder} distribution $q$ mapping from data space to latent space for the signals up to and including timestep $t_m$. Combined, we model the full distribution $p(\mathbf{x}, \mathbf{z})$ using sequences of latent trajectories to represent sequences of signals in data space.

In addition to representing a signal time series as a trajectory of a latent state, we aim to shape latent variable $\mathbf{z}$ with additional desiderata. For interpretability reasons we set $d=2$, i.e. a 2-dimensional latent space, to allow for visualization of the entire latent space at once; higher dimensionalities will be studied in future works. To allow for differentiating distinct regimes in the latent state, we require $\mathbf{z}$ to be structured for clustering:
\begin{align}
\mathcal{C}: \mathbf{z} \in \mathbb{R}^{d} \rightarrow \{1, \ldots, K\},
\end{align}
where $\mathcal{C}$ denotes a mapping of the latent variable to one of $K$ latent clusters. 

Finally, variable $\mathbf{z}$ should be informative w.r.t.\ the plasma's proximity to disruptions. We introduce a variable indicating this notion as an average risk of disruption: 
\begin{align}
    {D}_{\text{risk}} : \mathbf{z} \in \mathbb{R}^{d} \rightarrow [0, 1].
\end{align}
To quantify ${D}_{\text{risk}}$, we aim for it to represent the fraction of shots that eventually disrupt\footnote{Since this quantity is defined using the organization of plasma regimes by a learned, unknown latent variable, we do not learn the rate directly in a supervised setting. However, post-hoc analysis can be done on learned variable ${D}_{\text{risk}}$ to calibrate it and to quantify whether it can accurately represent the disruption rate.}. We can consequently interpret it as a continuous analogue of the disruption rate, where rather than selecting groups of shots a priori, we exploit the smooth manifold of learned latent variable $\mathbf{z}$ to automatically define the disruption rate for different parameter spaces. 

Note that we do not model explicitly how close in time we are to a disruption. Since we aim to model the entire operational space rather than the chain-of-events inevitably leading to a disruption, and given that we do not take into account future control actions, predicting a future disruption is ill-posed. Rather, ${D}_{\text{risk}}$ is used to organize the latent space into different regions associated with disruptions to find common patterns and to provide an average level of risk. It is also related, but not equivalent, to the notion of disruptivity, defined as the number of disruptions per second for a given plasma property space~\cite{deVries2011,disruptionpredictionITPA}. We can project a continuous equivalent of the disruptivity onto the latent space, which we denote as $\hat{D}_{\text{disr}}$. We additionally compare ${D}_{\text{risk}}$ to $\hat{D}_{\text{disr}}$.

\section{Dataset}\label{sec:data}
\begin{figure}[t]
\begin{center}\includegraphics[width=0.95\linewidth]{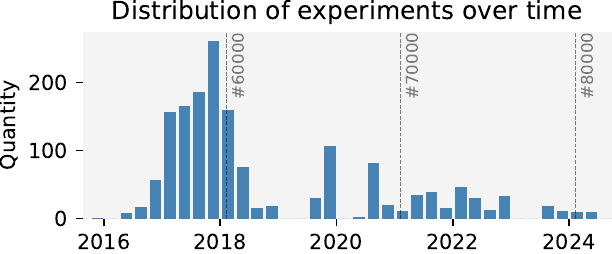}\end{center}
    \caption{The distribution of the discharges' dates, binned per quartile. Discharges range from TCV \#51325 to \#81751, with the majority taking place between 2016-2019.}
    \label{fig:dataset_when_2}%
\end{figure}

\begin{figure}[t]
\begin{center}\includegraphics[width=1\linewidth]{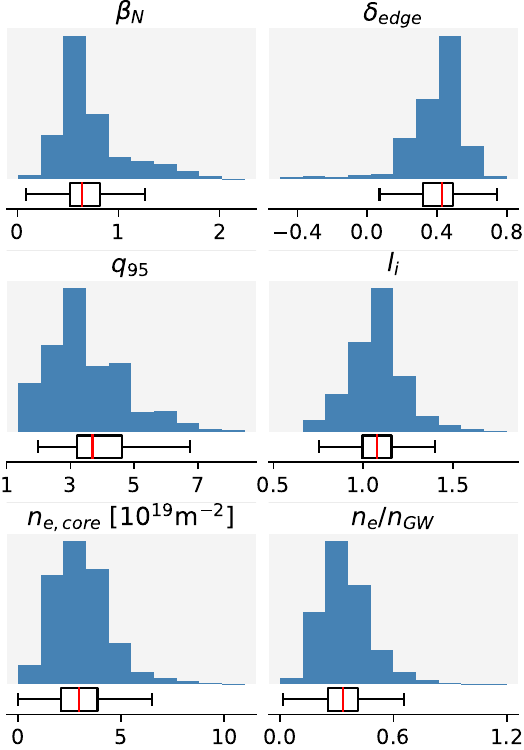}\end{center}
    \caption{Distributions of key plasma parameters in the dataset. The plotted values are averages over phases of \SI{20}{\milli\second}--around the TCV energy confinement time--to exclude transient states.}
    \label{fig:dataset_eda_2}%
\end{figure}

\begin{table*}[h]
{\renewcommand{\arraystretch}{0.97}
\begin{tabular}{p{0.14\textwidth}p{0.09\textwidth}p{0.71\textwidth}}
\toprule
\noalign{\vskip -0.4ex}
Variable & Unit & Description \\
\noalign{\vskip -0.4ex}
\midrule\arrayrulecolor[rgb]{0.9, 0.9, 0.9}
$A_p$ & \SI{}{\meter\squared} & Plasma cross-sectional area \\
$\Delta_{\text{min}}$ & \SI{}{\meter} & Minimum radial gap between the plasma edge and the inner or outer wall\\
$\delta_{\text{bottom}}$ & \hphantom{.} & Lower (bottom) plasma triangularity \\
$\delta_{\text{top}}$ & \hphantom{.} & Upper (top) plasma triangularity \\
$\kappa$ & \hphantom{.} & Plasma elongation \\\hline
$\text{AXUV}_{\text{X-point}}$ & $\text{a.u.}$ & Emissions measured using AXUV diodes covering the X-point region \\
$\text{PD}^{\textit{CIII}}_{\text{FFT}}$ & $\text{a.u.}$ & Spectral distribution of photodiode signal for CIII line emission ($\lambda$=465.1 nm), computed as the variance of frequency spectra for windows of $\SI{20}{\milli\second}$ \\\hline
$I_{p}$ & $\SI{}{\ampere}$ & Plasma current \\
$l_i$ & \hphantom{.} & Internal inductance of the plasma current \\
$q_{95}$ & \hphantom{.} & Safety factor at 95\% of enclosed magnetic flux \\\hline
$n_{e,\text{core}}$ & \SI{}{\per\meter\squared} & Vertical interferometer line-integrated electron density from 0.87 m < ch < 0.91 m \\
$n_e/n_{\textit{GW}}$ & \hphantom{.} & Greenwald fraction~\cite{Greenwald2002} using electron density measurements from interferometry \\\hline
$\textit{SXR}_{\text{core}}$ & \SI{}{\watt\per\meter} & Soft X-Ray core ($\rho_{\psi} < 0.15$) emission \\\hline
$P_{\textit{in}}$ & $\SI{}{\watt}$ & Total input power \\\hline
${W}_{\textit{tot}}$ & $\SI{}{\joule}$ & Total plasma stored energy \\
$\beta_{N}$ & \hphantom{.} & Normalized toroidal beta ($100\cdot\beta_{t}\frac{aB_0}{I_p \text{ [MA]}} $)\\\hline
$\textit{LM}$ & \SI{}{\tesla} & Locked mode amplitude \\
$\textit{RMS}_{(n=1)}$ & \SI{}{\tesla} & Root-mean-square of the $n=1$ mode amplitude for windows of \SI{2}{\milli\second} \\
$\textit{RMS}_{(n=2)}$ & \SI{}{\tesla} & Root-mean-square of the $n=2$ mode amplitude for windows of \SI{2}{\milli\second} \\
$\gamma_{\text{VGR}}$ & \SI{}{\hertz} & Vertical growth rate estimate using RZIp~\cite{marchioni2024} \\
\arrayrulecolor[rgb]{0., 0., 0.}
\bottomrule
\end{tabular}

\caption{The list of input signals and constructed features used in the latent variable model. Equilibrium-related features originate from LIUQE~\cite{moret2015liuqe}, MHD markers are computed using fast magnetic probes~\cite{testa2020}.}\label{tab:signals_2}
}
\end{table*}

We utilize a dataset of approximately 1600 TCV discharges, ranging from TCV \#51325 (Dec 2015) to \#81751 (Jun 2024), see Figure~\ref{fig:dataset_when_2} for the distribution of shots over time. To illustrate the variety of plasma scenarios we plot the distribution of key parameters in Figure~\ref{fig:dataset_eda_2}.

\subsection{Dataset construction}
The aim of this work is to characterize plasma dynamics in the flat-top phase of the discharge. The ramp-up and ramp-down phases are not considered here due to their non-stationary nature and distinct plasma trajectories, and are left for future investigation. To cover a representative sample of the TCV flat-top operational space, we construct the dataset by sampling thousands of past experiments and then filter on discharges that either (a) disrupt in the flat-top phase or (b) have a regular termination. Additionally, we constrain the scope to plasmas that reach a lower single null (X-point) configuration at some point during the flat-top phase. The dataset statistics are summarized as follows:
\begin{tcolorbox}[breakable,colback=bgcolor,colframe=bgcolor,width=\linewidth,enlarge left by=-0mm,arc = 0pt,outer arc=0pt,boxsep=-1mm, left=0mm, right=4mm, top=1mm, bottom=1mm,title=]
\begin{itemize}[itemsep=1pt]
    \item \SI{1629}{} discharges ranging from \#51325 to \#81751
    \item \SI{1768.96}{\second} of flat-top plasma dynamics
    \begin{itemize}\item[$\circ$]On average, $\SI{1.086}{\second} \pm \SI{0.49}{\second}$ per shot
    \item[$\circ$]\SI{297.53}{\second} with limited, \SI{1471.43}{\second} with diverted configuration\end{itemize}
    \item \SI{1147}{} flat-top disruptions, \SI{482}{} regular terminations
    \begin{itemize}\item[$\circ$]Disruption rate of \SI{70.4}{\percent}\footnotemark\end{itemize}
\end{itemize}
\end{tcolorbox}
\footnotetext{Note that the dataset is biased towards disruptive plasmas, and 70.4\% does not necessarily reflect the disruption rate of TCV operation.}

The dataset consists of a time series for each shot, where features correspond to diagnostic measurements or features derived from them, see Subsection~\ref{ss:disrsignals} for more details. Additionally, it contains disruption-related metadata, such as the time of disruption $t_D$. We define $t_D$ as the onset of the thermal quench leading to the eventual disruption, and automatically compute these times using the DEFUSE framework~\cite{defuseiaea}. As extra metadata used for the evaluation of the identified latent space, we leverage event detectors from the DEFUSE framework to populate the database\footnote{The development and evaluation of the automated event detection in DEFUSE is still ongoing. Nevertheless, early evaluations indicate that the aggregate distributions of detections are statistically meaningful.}. For example, events related to MHD (magnetohydrodynamics) markers for rotating/locked modes or indicators related to abnormal kinetic profiles (e.g. peaking or hollowing thereof). Finally, we leverage automated tools for confinement state classification~\cite{poels2025ldh} for evaluation, to correlate the identified states with different confinement modes. Specifically, we use the full ensemble presented in~\cite{poels2025ldh} to predict all confinement states and use detections with at least 75\% prediction confidence. These predictions cover \SI{1533.84}{\second} of plasma dynamics: \SI{1287.18}{\second} L-mode (83.9\%), \SI{16.56}{\second} of dithering (1.1\%) and \SI{230.10}{\second} H-mode (15.0\%).

\subsection{Signals}\label{ss:disrsignals}
The utilized signals are selected either because they are representative of plasma scenarios and/or because of their connection to known operational limits. An overview of all signals is provided in Table~\ref{tab:signals_2}. The signals are interpolated to a common timebase of \SI{10}{\kilo\hertz} using the last available sample---we refrain from non-causal interpolation methods (e.g., linear, spline) to ensure no information leakage about future plasma dynamics. Data retrieval is handled through the DEFUSE framework~\cite{defuseiaea}.

\section{Method}\label{sec:method}
The approach consists of a latent variable model that represents time series of discharge measurements as trajectories in a latent space. This latent space should be structured to distinguish different operating regimes, while being connected to disruption characteristics. We extend the Variational Autoencoder (VAE) framework~\cite{Kingma2014,rezende2014}, taking into account aforementioned desiderata. We introduce the VAE framework in Section~\ref{ss:vae_intro}, our structure in Section~\ref{ss:mmvae}, the training and inference procedure in Section~\ref{ss:traininferencevae}, followed by details on the architecture in Section~\ref{ss:vaearch}.

\subsection{The variational autoencoder}\label{ss:vae_intro}
\begin{figure*}[t]
\begin{center}\includegraphics[width=0.7\linewidth]{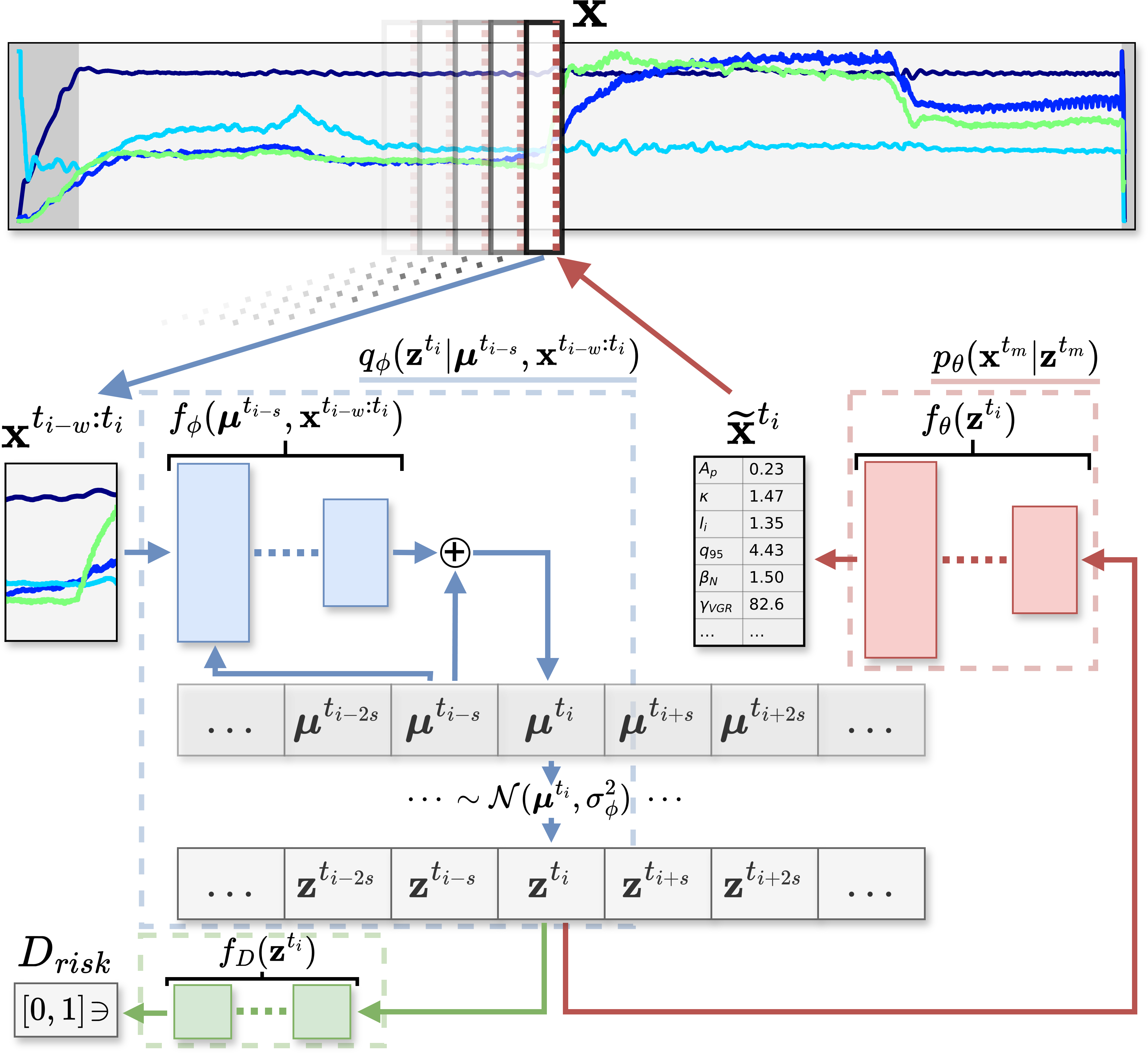}\end{center}
    \caption{Schematic overview of the model structure. The model consists of encoder distribution $q_\phi(\mathbf{z}^{t_i}|\boldsymbol{\mu}^{t_{i-s}}, \mathbf{x}^{t_{i-w}:t_i})$ (blue), decoder distribution $p_\theta(\mathbf{x}^{t_i}|\mathbf{z}^{t_i})$ (red) and disruption risk map $D_{\textit{risk}}(\mathbf{z}^{t_i})$ (green). Signals are encoded using timewindows of input signals, with the location in the latent space being computed as an update w.r.t.\ the previous location. Conversely, the mapping from latent space to data space and to the disruption risk are static in time.}
    \label{fig:schematic}%
\end{figure*}
The Variational Autoencoder is a generative model that aims to model a data distribution $p(\mathbf{x})$ under the assumption that it is generated using latent variable $\mathbf{z}$. We consequently model $p(\mathbf{x})$ as $p(\mathbf{x}, \mathbf{z}) = p(\mathbf{x}|\mathbf{z})p(\mathbf{z})$, choose prior distribution $p(\mathbf{z})$ and learn the data distribution as $p(\mathbf{x}|\mathbf{z})$. In the VAE framework, we additionally learn approximate posterior distribution $q(\mathbf{z}|\mathbf{x})$. Distributions $q(\mathbf{z}|\mathbf{x})$ and $p(\mathbf{x}|\mathbf{z})$ are often referred to as the encoder and decoder, respectively, given that they encode/decode from data space to latent space and vice versa. 

The encoder and decoder distribution in a VAE are parametrized by neural networks (NNs). In the standard formulation, the encoder is parametrized as follows:
\begin{align}\label{eq:qvae}
    q_\phi(\mathbf{z}|\mathbf{x}) &= \mathcal{N}(\mathbf{z}; \boldsymbol{\mu}_\phi(\mathbf{x}),\boldsymbol{\sigma}^2_\phi(\mathbf{x})I),\\
    \boldsymbol{\mu}_\phi(\mathbf{x}),\boldsymbol{\sigma}_\phi(\mathbf{x}) &= f_\phi(\mathbf{x}),
\end{align}
for model parameters $\phi$. For a Gaussian decoder, we can similarly parametrize the mean and variance. However, in practice, we often only parametrize the mean and fix the variance term:
\begin{align}
    p_\theta(\mathbf{x}|\mathbf{z}) = \mathcal{N}(\mathbf{x}; \boldsymbol{\mu}_{\theta}(\mathbf{z}), \sigma_\theta^2 I), \\
    \boldsymbol{\mu}_{\theta}(\mathbf{z}) = f_\theta(\mathbf{z}),
\end{align}
for model parameters $\theta$. Both the encoder and decoder are jointly optimized using a lower bound on the log likelihood of the data, the evidence lower bound (ELBO):
\begin{align}\label{eq:elbo}
  \text{ELBO}:= \quad \ln p(\mathbf{x}) \geq 
   &\underbrace{\mathbb{E}_{\mathbf{z}\sim q_\phi(\mathbf{z}|\mathbf{x})}[\ln p_{\theta}(\mathbf{x}|\mathbf{z})]}_{-\mathcal{L}_{\textit{rec}}(\mathbf{x})}
   \\-&\underbrace{\mathbb{E}[\ln q_\phi(\mathbf{z}|\mathbf{x}) - \ln p(\mathbf{z})]}_{{\mathcal{L}_{\textit{KL}}(\mathbf{z})}}.
\end{align}
The ELBO term can be split into two parts. The former, $\mathcal{L}_{\textit{rec}}$, can be considered as a reconstruction error. Usually, the expectation terms are approximated by sampling, which corresponds to taking data points and encoding and decoding them. Then, the error between the input samples and the reconstructions is minimized. The latter, $\mathcal{L}_{\textit{KL}}$, can be considered a regularization term. It is equivalent to the Kullback–Leibler divergence~\cite{kl1951}, and expresses how well the learned latent distribution is covered by the prior distribution. Often, it is approximated by taking samples of $\mathbf{z}$ and computing the relative log-likelihoods.

In order to get stable gradient estimates while training, latent variable $\mathbf{z}$ in Equation~\ref{eq:qvae} is sampled using the reparameterization trick~\cite{Kingma2014}. We parametrize parameters $\boldsymbol{\mu}$ and $\boldsymbol{\sigma}$ of the distribution and sample using external noise $\boldsymbol{\epsilon} \sim \mathcal{N}(\mathbf{0}, \mathbf{I})$ as $\mathbf{z} = \boldsymbol{\mu} + \boldsymbol{\sigma} \odot \boldsymbol{\epsilon}$ (where $\odot$ denotes element-wise multiplication). 

\begin{figure*}[t]
\begin{center}\includegraphics[width=0.96\linewidth]{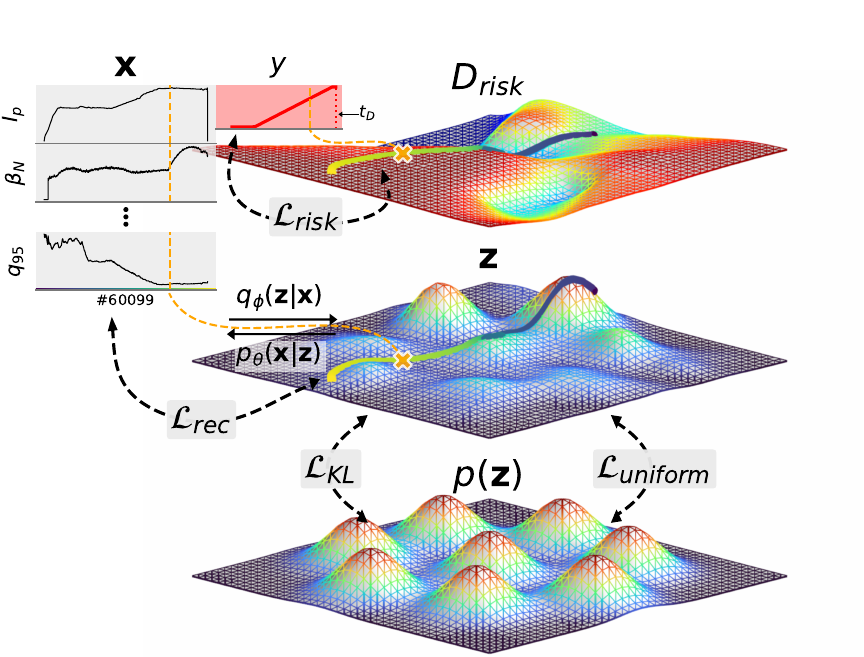}\end{center}
    \caption{Depiction of the training and inference procedure of the proposed method. Data $\mathbf{x}$, time series of signals corresponding to TCV experiments, are projected to latent variable $\mathbf{z}$ using approximate posterior, the encoder, $q_\phi(\mathbf{z}^{t_m}|\boldsymbol{\mu}^{t_{m-s}},\mathbf{x}^{t_{m-w}:t_m})$ (or $q_\phi(\mathbf{z}|\mathbf{x})$ aggregating over time). The generative distribution, the decoder, $p_{\theta}(\mathbf{x}^{t_m}|\mathbf{z}^{t_m})$ (or $p_\theta(\mathbf{x}|\mathbf{z})$ aggregating over time) provides the map back to data space. A timeslice in data space corresponds to a timeslice in latent space, consequently projecting discharges as latent trajectories. Simultaneously, we learn disruption risk map $D_{\textit{risk}}$ as a function of $\mathbf{z}$, using proxy labels $y$. The latent space is optimized to maximize the data likelihood ($\mathcal{L}_\textit{rec}$), match disruption information ($\mathcal{L}_{\textit{risk}}$), minimize divergence to prior modes ($\mathcal{L}_{\textit{KL}}$) while covering all modes of said prior ($\mathcal{L}_\textit{uniform}$).}
    \label{fig:multimodalvequentialvae}%
\end{figure*}

\subsection{Sequential multimodal VAE with disruption risk}\label{ss:mmvae}
To better address the problem statement, we extend the VAE formulation to incorporate a notion of time, to model the disruption risk, and introduce a structure for clustering. We first introduce the extended formulation for all components, including the neural network functions we learn in practice. Then, we describe the prior distribution and its role in clustering in the latent space.

\textbf{Components.} Similar to recurrent VAEs~\cite{Girin2021,recurrentvae2020}, we introduce the notion of time into the VAE by modeling a dependence between the latent variable at a given timestep, $\mathbf{z}^{t_m}$, and its past states, $\mathbf{z}^{\mathbf{t} < t_m}$, see also Equation~\ref{eq:encproblemstatement}. However, for efficiency reasons, we do not learn a dependence on all signals at each timestep. Rather, we operate using a fixed timewindow of size $w$, which we slide across input signals with stride $s$. Then, our encoding distribution is reformulated as follows:
\begin{align}
    \mathcal{T}_i = \{0, s, 2s, \ldots, m\}&,\\
    \begin{split}q(\mathbf{z}^{t_1}, \ldots, \mathbf{z}^{t_m}|\mathbf{z}^{t_0},\mathbf{x}^{\mathbf{t}\leq{t_m}})= \ \ \ \ &\\ \prod_{i\in \mathcal{T}_i}q(\mathbf{z}^{t_i}|\mathbf{z}^{t_{i-s}}, \mathbf{x}^{t_{i-w}:t_i})&,\end{split}
    \label{eq:encoderfinal}
\end{align}
assuming $m$ as a multiple of $s$. That is, we model sequences at a sampling rate of $\frac{1}{s}$ using as input timewindows of signals of size $w$ and the previous latent state. For the decoder, we model a single timestep, i.e.,
\begin{align}
    p_{\theta}(\mathbf{x}^{t_m}|\mathbf{z}^{t_m}),
\end{align}
for the plasma discharge at time $t_m$. The fixed mapping allows for interpretation of the latent variable by projecting it back directly to physics quantities in data space.

Both the encoder and decoder are parametrized by neural networks as follows. To encourage continuity in the latent space, we formulate the encoder distribution to update the location with respect to the previous timestep. To reduce noise while training, we utilize the distribution mean directly rather than sampling at each timestep in the trajectory, i.e., we approximate $q$ as follows:
\begin{align}
    \hspace{-.25cm}q(\mathbf{z}^{t_m}|\mathbf{z}^{t_{m-s}}, \mathbf{x}^{t_{m-w}:t_m}) \approx q_\phi(\mathbf{z}^{t_m}|\boldsymbol{\mu}^{t_{m-s}}, \mathbf{x}^{t_{m-w}:t_m}),\hspace{-.1cm}
\end{align}
for parameters $\phi$, where $\boldsymbol{\mu}^{t_{i-s}}$ is the mean of $q_\phi$ at the previous timestep. Then, the structure becomes:
\begin{align}
\begin{split}
    q_\phi(\mathbf{z}^{t_m}|\boldsymbol{\mu}^{t_{m-s}}, \mathbf{x}^{t_{m-w}:t_m}) &= \\\mathcal{N}(\mathbf{z}^{t_m}; \boldsymbol{\mu}^{t_{m-s}} +\mu_\phi(&\boldsymbol{\mu}^{t_{m-s}},\mathbf{x}^{t_{m-w}:t_m}),\sigma_\phi^2I),
    \end{split}\\
    \mu_\phi(\boldsymbol{\mu}^{t_{m-s}},\mathbf{x}^{t_{m-w}:t_m}) &= f_\phi(\boldsymbol{\mu}^{t_{m-s}}, \mathbf{x}^{t_{m-w}:t_m}),\label{eq:encoderfullmodel}
\end{align}
for neural network function $f_\phi$. For stability during training~\cite{dang2024beyond} and fast projections of large sets of data distributions, we fix $\sigma_\phi$ as a model hyperparameter. Note that we parametrize the mean of $q_\phi$ using the previous mean summed to the neural network output. That is, we can consider the network as a coarse neural differential equation~\cite{chen2018} that models the mean of the latent distribution, 
which is subsequently solved using a forward Euler scheme with fixed timestep $s$. 

For the decoder distribution we parameterize the mean and fix the variance:
\begin{align}
    p_\theta(\mathbf{x}^{t_m}|\mathbf{z}^{t_m}) = \mathcal{N}(\mathbf{x}^{t_m}; \boldsymbol{\mu}_{\theta}(\mathbf{z}^{t_m}), \sigma_\theta^2 I), \\
    \boldsymbol{\mu}_{\theta}(\mathbf{z}^{t_m}) = f_\theta(\mathbf{z}^{t_m}),
\end{align}
for neural network function $f_\theta$. Additionally, we learn a disruption variable ${D}_{\text{risk}}$ as a function of $\mathbf{z}$:
\begin{align}
    D_{\textit{risk}}(\mathbf{z}^{t_m}) = f_D(\mathbf{z}^{t_m}),
\end{align}
for neural network function $f_D$. An overview of the model components and their interaction is given in Figure~\ref{fig:schematic}.

\textbf{Multimodality.} To introduce a structure suited for identifying different regimes, we formulate prior distribution $p(\mathbf{z})$ as a multimodal distribution. Specifically, we use a mixture of Gaussians:
\begin{align}\label{eq:vaeprior}
        p(\mathbf{z}) &= \sum_{k=1}^K w_k \mathcal{N}(\mathbf{z};\boldsymbol{\mu}_{k}, \sigma_k^2),
\end{align}
for a mixture of $K$ Gaussian distributions parameterized by means $\boldsymbol{\mu}_k$, standard deviations $\boldsymbol{\sigma}_k$, and mixing weights $w_k$. Assuming sufficient distance between the distributions, $p(\mathbf{z})$ will have multimodal structure, encouraging our learned latent variable to similarly have multiple peaks.

For simplicity, we fix the standard deviation for all distributions, and choose equal mixing weights $w_k = \frac{1}{K}$. Then, we can define the likelihood of being assigned to a cluster as the likelihood under the Gaussian mixture:
\begin{align}\label{eq:clustering_t}
    \mathcal{C}:= p(\mathcal{C}=k|\mathbf{z}) 
= \frac{\mathcal{N}(\mathbf{z}|\boldsymbol{\mu}_k, \sigma_p^2)}{\sum_{j=1}^{K} \mathcal{N}(\mathbf{z}|\boldsymbol{\mu}_j,\sigma_p^2)},
\end{align}
with fixed prior standard deviation $\sigma_p$.

\subsection{Training and inference}\label{ss:traininferencevae}
To train NN functions $f_\phi$ (encoder), $f_\theta$ (decoder) and $f_D$ (disruption risk), we extend the ELBO. Given an input timewindow of signals $\mathbf{x}^{t_{m-w}:t_m}$, we sample the latent variable:
\begin{align}
\mathbf{z}^{t_m} &\sim q_\phi(\mathbf{z}^{t_m}|\boldsymbol{\mu}^{t_{m-s}},\mathbf{x}^{t_{m-w}:t_m}),
\end{align}
whereas for the reconstruction, we directly use the mean of the decoder:
\begin{align}
    \mathbf{\widetilde{x}}^{t_m} = \boldsymbol{\mu}_\theta(\mathbf{z}^{t_m}).
\end{align}
Then, the reconstruction reduces to a squared error as follows:
\begin{align}
    \mathcal{L}_{\textit{rec}}(\mathbf{x}^{t_m}) &= (\mathbf{\widetilde{x}}^{t_m} - \mathbf{x}^{t_m})^2. 
\end{align}
The regularization term is as follows:
\begin{align}
    \mathcal{L}_{\textit{KL}}(\mathbf{z}) &= \textit{KL}\infdivx{q_\phi(\mathbf{z}^{t_m}|\mathbf{z}^{t_{m-s}},\mathbf{x}^{t_{m-w}:t_m})}{p(\mathbf{z})},
\end{align}
which we estimate by using samples of $\mathbf{z}^{t_m}$ and computing the relative log-likelihood under both the prior and posterior. Since the KL divergence can act mode seeking in this setting~\cite{minka2005divergence,klmodeseeking}, we utilize it to encourage single samples of $\mathbf{z}$ to lie on one mode of the prior $p(\mathbf{z})$, and dynamically set its mixing weights $w_k$ as $p(C = k|\mathbf{z})$ (Equation~\ref{eq:clustering_t}). To encourage coverage of all modes, given that we assume equal mixing weights for the full distribution, we introduce a separate loss term $\mathcal{L}_{\textit{uniform}}$. For a batch of latent samples $\{\mathbf{z}^{(n)}\}_{n=1}^{N}$, we can define the average inverse distance to a component of the prior distribution as follows:
\begin{align}
    \bar{d}_k(\mathbf{z}) 
= \frac{1}{N}\sum_{n=1}^{N}\frac{1}{\|\mathbf{z}^{(n)} - \boldsymbol{\mu}_k\|^2}.
\end{align}
If the prior components are sufficiently far apart and we utilize equal mixing weights, the inverse distances $\bar{d}_k(\mathbf{z}) \in K$ approximate a uniform distribution for samples of prior $p(\mathbf{z})$. Consequently, we use this property to optimize our latent space to cover all models equally in expectation\footnote{One can draw a parallel between this formulation and learning latent variables that are uninformative w.r.t.\ a property of choice, e.g.~\cite{zheng2019}.}:
\begin{align}
\begin{split}
    \mathbb{E}_{\mathbf{z}\sim q_\phi(\mathbf{z|x})}&\left[\text{softmax}\left([\bar{d}_1(\mathbf{z}),\,\dots,\,\bar{d}_K(\mathbf{z})]\right)\right]
= \\&\hspace{3.5cm}\left[\frac{1}{K},\,\dots,\,\frac{1}{K}\right],
\end{split}
\end{align}
with short notation of the posterior distribution and its samples for readability. To optimize this expectation, we minimize the negative log-likelihood through a cross-entropy loss term:
\begin{align}
\begin{split}
&\mathcal{L}_{\textit{uniform}}(\mathbf{z}) 
=\\ -&\sum_{k=1}^{K}\frac{1}{K}\log\left(\text{softmax}\left([\bar{d}_1(\mathbf{z}),\,\dots,\,\bar{d}_K(\mathbf{z})]\right)_k\right).
\end{split}
\end{align}

For the disruption risk variable, we create labels $y^{t_m}$ using the time of disruption $t_D$ as computed in~\cite{defuseiaea}. If a discharge did not disrupt, it is set to 0 for each timestep. For discharges that disrupted, we define $y^{t_m}$ to be 1 shortly before $t_D$ with a linear ramp leading up to it as follows:
\begin{equation}\label{eq:disrlabel}
    \hspace{-0.1cm}y^{t_m} =
\begin{cases}
    1, & \text{if } t_m \geq t_D - A \\
    \frac{(t_m - (t_D - B))}{(B - A)}, & \text{if } t_D - B \leq t_m < t_D - A \\
    0, & \text{otherwise,}
\end{cases}
\end{equation}
for hyperparameters $A$ and $B$. Then, we optimize the disruption risk using binary cross-entropy: 
\begin{align}
\begin{split}
    \mathcal{L}_{\textit{risk}}(\mathbf{z}) =
    -\big[\,y^{t_m}\log(D_{\textit{risk}}(\mathbf{z}^{t_m})) \\+\ (1 - y^{t_m})\log(1 - D_{\textit{risk}}(\mathbf{z}^{t_m}))\,\big].
\end{split}
\end{align}
The total loss of the model is the sum of all components:
\begin{align}\label{eq:totalloss}
\mathcal{L} = a \mathcal{L}_{\textit{rec}} + b \mathcal{L}_{\textit{KL}} + c\mathcal{L}_{\textit{uniform}} + d\mathcal{L}_{\textit{risk}},
\end{align}
for loss weights $a$, $b$, $c$ and $d$. We train using minibatches of sequences from different shots. The sequences are split into the respective strided timewindows, and iteratively parsed by the model components. All parameters are jointly optimized with gradient-based optimization, that is, we apply backpropagation through time~\cite{werbos1988}. Remaining details are provided in~\ref{ap:vaeparameters}. An overview of the training and inference setup is given in Figure~\ref{fig:multimodalvequentialvae}. 

\subsection{Architecture}\label{ss:vaearch}
Finally, we summarize the implementation of neural network functions $f_\phi$, $f_\theta$ and $f_D$. For the encoder $f_\phi(\boldsymbol{\mu}^{t_{m-s}}, \mathbf{x}^{t_{m-w}:t_m})$ we utilize the Fourier Neural Operator (FNO)~\cite{li2021} to extract temporal information from timewindow input $\mathbf{x}^{t_{m-w}:t_m}$, as it has shown strong performance on modeling various fusion-related time series~\cite{Gopakumar2024PlasmaSurrogate,Poels2023neuralPDE,gopakumar2025calibrated,pamela2025neural,poels2025ldh}. The FNO performs nonlinear, global convolutions by applying linear global transformations in the frequency domain and nonlinear local transformations in the temporal domain. An FNO layer can be denoted as follows:
\begin{align}
    \hspace{-.2cm}\textit{FNO}^i: \mathbf{h}^i = \psi\big(\text{FFT}^{-1}(\mathbf{R}^i\text{FFT}(\mathbf{h}^{i-1})) + \mathbf{W}^i \mathbf{h}^{i-1}\big),
    \label{eq:fno}
\end{align}
which maps an input signal $\mathbf{h}^{i-1}$ to its output $\mathbf{h}^i$. $\mathbf{R}^i \in \mathbb{R}^{D \times D \times M}$ and $\mathbf{W}^i \in \mathbb{R}^{D \times D}$ ($D$ hidden dimensions; $M$ fourier modes) are the learned parameters, $\psi$ denotes the nonlinear activation function, and FFT denotes the Fast Fourier Transform~\cite{cooley1965}.

The output of the FNO layers is flattened and concatenated to the previous value of $\boldsymbol{\mu}$, which is subsequently put through a small Multi-Layer Perceptron (MLP)~\cite{ivakhnenko1971}. We can summarize the encoder as:
\begin{align}
f_\phi(\cdot) = \textit{MLP}_{\phi}(\textit{FNO}_{\phi}(\mathbf{x}^{t_{m-w}:t_m}),\boldsymbol{\mu}^{t_{m-s}}).
\end{align}

\begin{figure}[t]
\begin{center}\includegraphics[width=0.84\linewidth]{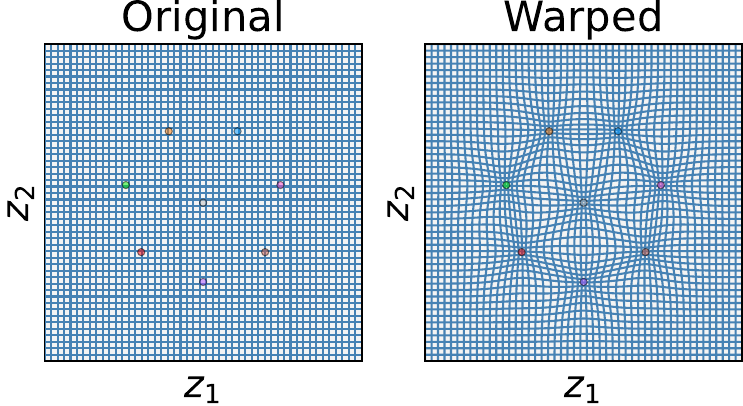}\end{center}
    \caption{Precomputed deformation of a uniform space, used to ease the task of placing appropriate probability density on the prior modes during model training.}
    \label{fig:warp}%
\end{figure}

The decoder $f_\theta$ and disruption risk $f_D$ follow a static formulation, and are implemented as MLPs. However, the low-frequency bias of neural networks~\cite{xu2019lowfreq} could limit the expressivity of the learned latent variable. To mitigate this issue, we first apply positional encoding to the latent variable before applying an MLP. This approach has been shown to aid in representing higher-frequency information~\cite{Mildenhall2021nerf}, helping us better shape the latent space. The applied frequency encoding can be denoted as follows:
\begin{align}
\begin{split}
    \gamma(p) = \big[\sin(2^0\pi p), \cos(2^0\pi p), \ldots, \\\sin(2^{L-1}\pi p), \cos(2^{L-1}\pi p)\big],
\end{split}
\end{align}
for applying $L$ frequencies to parameter $p$. We apply it separately to both dimensions. With $({z}^{t_m}_1,{z}^{t_m}_2) = \mathbf{z^{t_m}}$ denoting the two latent dimensions, we can summarize the decoder and disruption map as:
\begin{align}
    f_\theta(\cdot) &= \textit{MLP}_{\theta}(\gamma({z}^{t_m}_1),\gamma({z}^{t_m}_2)),\\
    f_D(\cdot) &= \textit{MLP}_{D}(\gamma({z}^{t_m}_1),\gamma({z}^{t_m}_2)).
\end{align}
Additionally, to aid the optimization w.r.t.\ matching the prior\footnote{We found that training could lead to degenerate solutions where several prior modes were `ignored'. A precomputed warping towards these modes results in a higher likelihood of placing some samples on a new peak, providing more informative gradients w.r.t.\ unexplored territories of the latent space.}, we add a precomputed warping of the latent space using the prior's parameters. Using mixture components ($\boldsymbol{\mu}_k, \boldsymbol{\sigma}_k$), we place more density near the modes at the start of the optimization procedure, see Figure~\ref{fig:warp} for a depiction. An inverse of this transformation is applied on decoding from the latent space, ensuring it has no adverse effects on the decoder or disruption risk expressivity.

Finally, in order to better approximate the disruption rate using the disruption risk variable $D_{\textit{risk}}$, we calibrate the variable post hoc. That is, after the model is trained, we can compute the actual disruption rate of all points in the learned latent space by projecting the dataset and evaluating the proportion of discharges that eventually disrupt for each location in $\mathbf{z}$. To ensure a continuous map, the projection is performed using the latent trajectories' distributions, that is, we sum the probability density functions of the inferred latent parameters. Then, we apply Platt scaling~\cite{platt1999probabilistic} to calibrate the prediction curve of $D_{\textit{risk}}$ to best represent the disruption rate.

\section{Experiments and Results}\label{sec:experiments}
In this section we evaluate the proposed method for learning a model that projects discharge measurements onto a lower-dimensional, interpretable latent variable. This latent variable is optimized to represent the measurements in an informative manner w.r.t.\ operational limits. Consequently, we evaluate (1) latent space properties in the context of disruption metrics; (2) the ability to separate distinct types of operational limits; and (3) using the model to identify patterns that can facilitate large-scale analyses. We assess (3) by doing a demonstrative study on parameters correlated with different types of disruptions.

A summary of the training details and model hyperparameters is provided in Section~\ref{ss:expdatasplit}. In Section~\ref{ss:explatentspace} we provide an analysis of the identified latent space and its relation to disruption metrics. Section~\ref{ss:expclusteringdisruptions} evaluates the ability to distinguish different types of disruptions and expands on the interpretability of the latent space w.r.t.\ physics quantities. Finally, Section~\ref{ss:expcounterfactual} evaluates the utility of the model for downstream tasks, demonstrated with a proof-of-concept automatic analysis of disruption causes.

\subsection{Dataset split and hyperparameters}\label{ss:expdatasplit}
\textbf{Dataset split.} The dataset (1629 shots) is split into a training set (1300 shots), validation set (165 shots) and test set (164 shots). The test and validation set are selected to be representative of the overall distribution while still being dissimilar to the train data. To select validation and test shots, we first compute a distance matrix of all shots using average values of operational parameters $I_{p}$, $B_0$, $q_{95}$, $l_i$, $n_{e,\text{core}}$, $\beta_{N}$, $W_{\textit{tot}}$, $P_{\textit{in}}$ and $P_{\textit{rad}}$ (see also~\cite{poels2025ldh} for additional definitions) and subsequently find 10 clusters using agglomerative hierarchical clustering~\cite{murtagh2012}. Then, we construct the test and validation set using a diversity maximization approach~\cite{chandra2001}. We sample from each cluster in proportion to its size in the full dataset. Specifically, we iteratively select the most distant shots to first construct the test set, and subsequently the validation set. As a result, we reserve a set of shots for the test and validation set that are both distant to the train shots while still covering a diverse parameter space.

Nevertheless, most evaluations also use the training data, given that we explore the identified latent variable $\mathbf{z}$ and its correspondence to disruption-related metadata not used during training; it is largely an unsupervised problem. The validation set is used during training to detect overfitting, whereas the test set is used to quantitatively evaluate disruption risk metrics on unseen data.

\textbf{Hyperparameters.} The main hyperparameters consider the dataset-related parameters and the prior structure. Remaining details on training and the model architecture are found in~\ref{ap:vaeparameters}. The models are implemented using PyTorch~\cite{paszke2019}, and we use net:cal~\cite{kuppers2020} for post hoc Platt scaling.

For data-related parameters, we use timewindows of $w=50$ timesteps and an equivalent stride of $s=50$ timesteps (Equation~\ref{eq:encoderfinal}). Given a data sampling rate of \SI{10}{\kilo\hertz}, we generate latent trajectories with a timestep of \SI{5}{\milli\second}. The disruption labels are computed using a ramp starting at $B=\SI{1}{\second}$ before $t_D$ up to $A=\SI{0.15}{\second}$ before the time of disruption (Equation~\ref{eq:disrlabel}). $A$ is similar to the current redistribution time on TCV~\cite{labit2024}, whereas $B$ is selected heuristically to cover most of the flat-top dynamics, see~\ref{ap:vaeparameters} for more discussion. Additionally, all signals are standardized by subtracting the mean and dividing by the standard deviation using statistics computed on the train set.

For the prior structure (Equation~\ref{eq:vaeprior}), we scan a set of configurations that place one peak in the center surrounded by a set of equally spaced modes around this center. To quantify the benefit of adding more modes, we compute the mutual information~\cite{mutualinformation} between the states and disruption-related event detections~\cite{defuseiaea,poels2025ldh}. At 7 surrounding peaks, for a total of $K=8$ Gaussians in the mixture prior, the benefit of adding more modes levels off, see also Figure~\ref{fig:mi_states}. Consequently, we use this configuration as prior distribution.

\subsection{Latent space and disruption metrics}\label{ss:explatentspace}
\begin{figure}[t]
\begin{center}\includegraphics[width=1\linewidth]{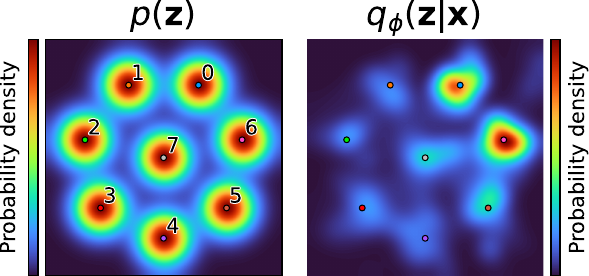}\end{center}
    \caption{The probability density of the chosen prior distribution $p(\mathbf{z})$ and the learned posterior distribution $q_\phi(\mathbf{z}|\mathbf{x})$. Peaks of the prior, corresponding to cluster mapping $\mathcal{C}$, are labeled 0-7 (left). The learned latent variable $q_\phi(\mathbf{z}|\mathbf{x})$ shows a clear multimodal structure as desired (right), with deviations from the prior due to the the competing objectives in the joint model optimization.}
    \label{fig:priorposterior}%
\end{figure}

\textbf{Latent space density.} Since we selected a dimensionality of 2 for latent variable $\mathbf{z}$, we can visualize the distribution in its entirety. We start with evaluating the spread around the latent space. In Figure~\ref{fig:priorposterior} (left) we plot the probability density of the selected prior with $K=8$ peaks, and label these peaks for the evaluation w.r.t.\ clusters later in this section. In Figure~\ref{fig:priorposterior} (right) we see the learned distribution given our dataset. The latent distribution still follows a multimodal structure as desired, with deviations from the prior because of competing optimization objectives. Note that since the variable is learned, the axes are arbitrary up to a scaling of the model settings. Consequently, for all projections on $\mathbf{z}$ in this section we omit axis labels but rather keep the domain fixed.

\begin{figure}[t]
\begin{center}\includegraphics[width=1\linewidth]{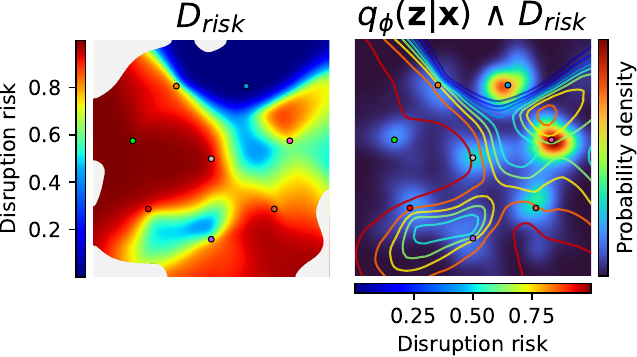}\end{center}
    \caption{The learned disruption risk variable $D_{\textit{risk}}$ (left), overlaid on the posterior distribution (right). Regions with low and high disruption risk are spread throughout the latent space, with a zone of risk-free plasmas projected on the blue peak in the top right.}
    \label{fig:disruptionprox}%
\end{figure}

\begin{figure}[t]
\begin{center}\includegraphics[width=1\linewidth]{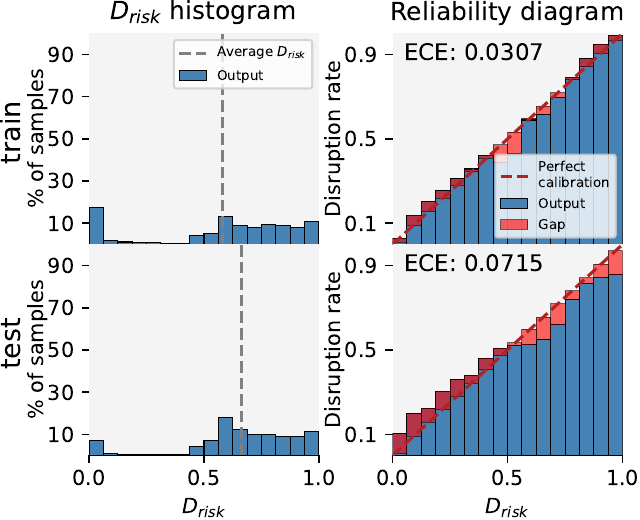}\end{center}
    \caption{The distribution (left) and reliability diagrams (right) of $D_{\textit{risk}}$ w.r.t.\ the disruption rate. The reliability diagram provides a visual depiction of the expected calibration error (see Equation~\ref{eq:ecedisr}), with the corresponding metric values plotted on top. We define the disruption rate as, for each location in $\mathbf{z}$, the fraction of timeslices projected there where the discharge ends in a disruption. 
    The value of $D_{\textit{risk}}$ is spread, with a notable lack of estimates between $\frac{1}{16}$ to $\frac{1}{2}$. It represents the actual disruption rate well, deviating only $\approx$3\% on average for shots used to identify the latent space (top), and $\approx$7\% for a set of novel discharges (bottom).}
    \label{fig:disrcalibration}%
\end{figure}

\begin{figure}[t]
\begin{center}\includegraphics[width=1\linewidth]{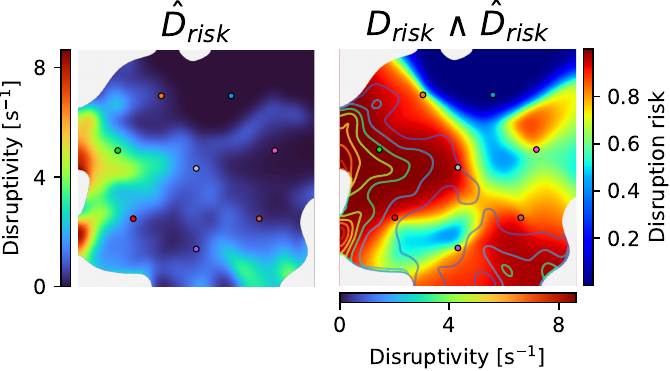}\vspace{3mm}\end{center}
    \caption{A projection of the disruptivity $\hat{D}_{\textit{disr}}$ (left), overlaid on the disruption risk ${D}_{\textit{risk}}$ (right). We define $\hat{D}_{\textit{disr}}$ as the number of disruptions per second for a given plasma property space. It is computed by summing the last projections of the disrupting discharges (0 to \SI{5}{\milli\second} before $t_D$) and dividing by posterior $q_\phi(\mathbf{z}|\mathbf{x})$, rescaled to the number of disruptions and the time span of the projections, respectively. }
    \label{fig:disruptivity}%
\end{figure}

\begin{figure}[t]
\begin{center}\includegraphics[width=0.75\linewidth]{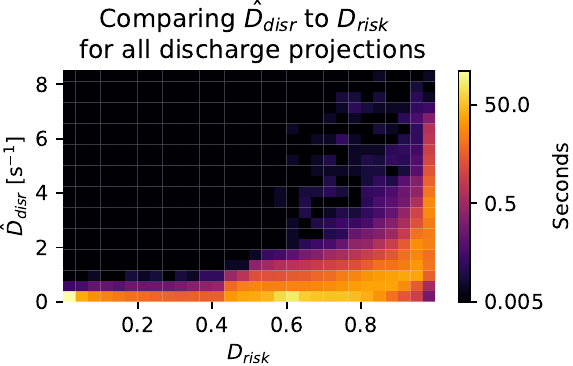}\end{center}
    \caption{A comparison of $D_{\textit{risk}}$ and $\hat{D}_{disr}$. For low estimates of disruption risk ($\leq 0.45$), we find near-0 rates of disruptivity. That is, plasma regimes with low estimates for $D_{\textit{risk}}$ are distant from the actual onset of disruptions. As the risk increases, the disruptivity increases accordingly, with an exponential-like curve.}
    \label{fig:disruptivityhist}%
\end{figure}

\textbf{Disruption risk.} The identified disruption risk variable $D_{\textit{risk}}$ is depicted in Figure~\ref{fig:disruptionprox}. The primary purpose of $D_{\textit{risk}}$ is as regularization that shapes the latent space to separate disruptive and nondisruptive regions. To validate that it can successfully provide such a separation, we fit a re-scaling of the `raw' model output using Platt scaling w.r.t.\ the disruption rate and evaluate the difference\footnote{The model did not have labels that represent the disruption rate during training. Consequently, directly comparing the raw model output to a calibrated quantity does not capture whether it represents the same information, rather, we fit a simple re-scaling to the model outputs first.}. We define this disruption rate as, for each location in $\mathbf{z}$, the fraction of timeslices projected there where the discharge ends in a disruption. We denote the real disruption rate as $\widetilde{D}_{rate}$ and describe the deviation to $D_{\textit{risk}}$ using the expected calibration error (ECE)~\cite{degroot1983,naeini2015}:
\begin{align}\label{eq:ecedisr}
    \hspace{-.11cm}\textit{ECE} = \sum_{m=1}^M \frac{N_m}{N}|\widetilde{D}_{\textit{rate}}(\mathbf{z}_{\mathbf{B}_m}) - D_{\textit{risk}}(\mathbf{z}_{\mathbf{B}_m})|,
\end{align}
with $N$ denoting the total number of samples. The ECE splits output range 0-1 into $M$ bins of $N_m$ samples each: the $m^\text{th}$ bin covers rate ${(\frac{m-1}{M}, \frac{m}{M}]}$. The samples falling in the respective bin are denoted with $\mathbf{z}_{\mathbf{B}_m}$, and the ECE consequently measures the deviation of the predicted rate to the real rate. We can interpret the ECE as the average error between $D_{\textit{risk}}$ and the real disruption rate.

The ECE and the accompanying reliability diagram---a visualization of the ECE---are provided in Figure~\ref{fig:disrcalibration}. $D_{\textit{risk}}$ accurately captures the real disruption rate, even when projecting a set of new, dissimilar shots (test set; bottom). Notably, there is a clear separation in the distribution of $D_{\textit{risk}}$ in the latent space: there are few locations with values between $\frac{1}{16}$ to $\frac{1}{2}$, rather, there is a region with almost no discharges ending in a disruption and a uniform spread for rates of $\approx$$\frac{1}{2}$ and higher.

\textbf{Disruptivity.} Next, we compare $D_{\textit{risk}}$ to the disruptivity $\hat{D}_{\text{disr}}$, defined as the number of disruptions per second for a given plasma property space~\cite{deVries2011,disruptionpredictionITPA}. We can compute this quantity in a continuous fashion by computing the distribution of projections just before $t_D$ and dividing by the posterior, rescaled to the number of disruptions and the number of seconds of plasma dynamics, respectively. Figure~\ref{fig:disruptivity} depicts $\hat{D}_{\text{disr}}$ and its comparison to $D_{\textit{risk}}$. Expectedly, they significantly overlap, with higher values of disruptivity in zones of values for $D_{\textit{risk}}$ approaching 1. Low values of $D_{\textit{risk}}$ still contain many discharges with regular terminations, consequently, they have not crossed an `uncontrollability' boundary. In Figure~\ref{fig:disruptivityhist} we plot the histogram comparing both quantities, better quantifying this notion. At low values of $D_{\textit{risk}}$, up to $\approx$0.45, we see virtually no disruptions, with an exponential curve as the risk increases. 

\begin{figure}[t]
\begin{center}\includegraphics[width=1\linewidth]{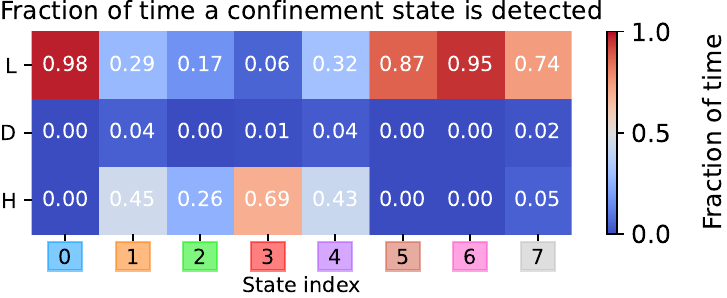}\end{center}
    \caption{Correlation between high-confidence ($\geq 0.75$) confinement state detections computed using~\cite{poels2025ldh} and the states found in $\mathbf{z}$. For each state we denote the fraction of time L, D or H-mode is detected for the total time spent in a state. Even though no confinement state labels are used in this work, some separation is recovered by virtue of clustering the operating regimes.}
    \label{fig:heatmapldh}%
\end{figure}

\textbf{Individual states.} Finally, we analyze correlations between the individual states, i.e., the peaks in the posterior distribution, and known plasma quantities. To do so, we compute the fraction of time an event is detected when a plasma is considered to be in a given state as defined by $\mathcal{C}$ (Equation~\ref{eq:clustering}). We first consider the plasma confinement state, automatically labeled using~\cite{poels2025ldh}. An overview of the fraction of time spent in each confinement state, for each state in $\mathbf{z}$, is provided in Figure~\ref{fig:heatmapldh}. Note that we only consider detections with $\geq$75\% confidence to ensure high-quality labels~\cite{poels2025ldh}, consequently, the detections do not necessarily sum to 1 for each state. We see a clear separation, with states 0, 5, 6 and 7 containing (almost) only L-mode detections, and mostly H-mode detections in the other states. Even when using no information about the confinement state for training, the latent space can distinguish these regimes. Additionally, we compute the correlations with disruption-related event detections from~\cite{defuseiaea}. We find less clear patterns compared to the confinement states, with the exception of neoclassical tearing modes (NTMs); see also Figure~\ref{fig:heatmap_events}. Most likely, many disruption-related events operate on much smaller timescales given the potentially very fast disruption dynamics in TCV~\cite{defuseiaea,marchioni2024,GTurri_2008}, whereas $\mathbf{z}$ represents dynamics on larger timescales connected to global operating regimes.

\begin{figure}[t]
\begin{center}\includegraphics[width=1\linewidth]{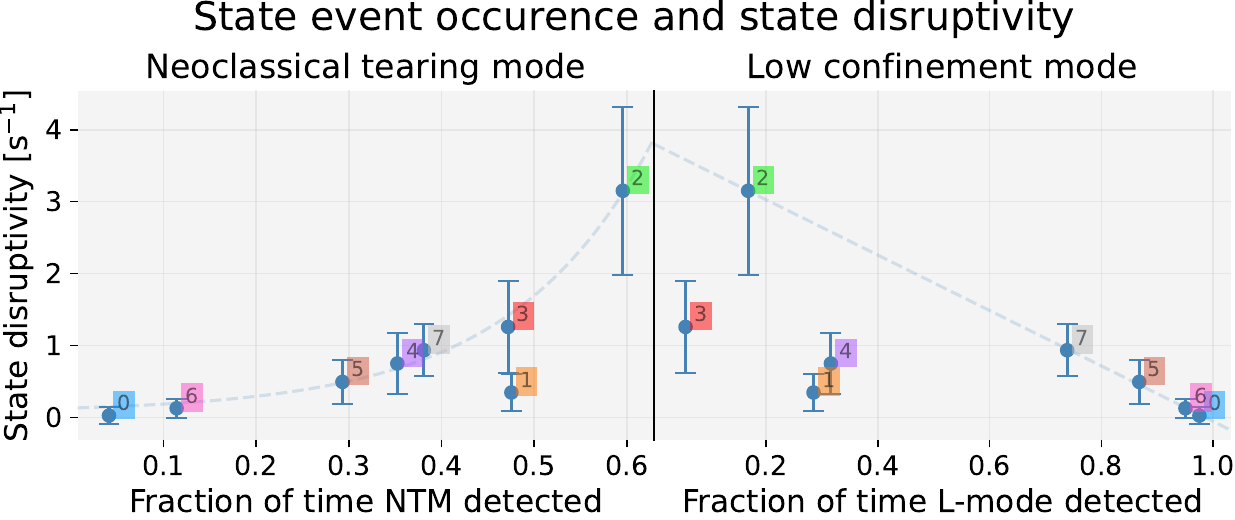}\end{center}
    \caption{Comparing the two most commonly detected events, NTMs and L-mode, to the mean disruptivity in each state ($\pm$ standard deviation). We see clear patterns, with NTM occurence being associated with more disruptions, and conversely L-mode detections  correlating with less disruptivity.}
    \label{fig:state_disruptivity}%
\end{figure}

Finally, we can correlate these events and their connection to the states' average disruptivity scores. For the two most commonly occuring detections, NTMs and L-mode, we plot this relation, see Figure~\ref{fig:state_disruptivity}. There is a clear correlation between the respective event occurrence and average disruptivity in each state, validiting the sensibility of the identifies states. For example, NTMs are often observed in connection to disruptions in TCV~\cite{labit2024}, and similarly L-mode plasmas generally operate further away from operational boundaries~\cite{deVries2011}.

\subsection{Distinguishing different types of 
disruptions}\label{ss:expclusteringdisruptions}
\begin{figure}[t]
\begin{center}\includegraphics[width=0.96\linewidth]{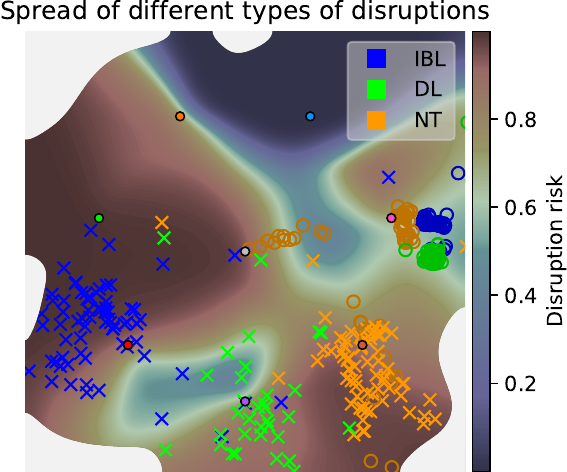}\end{center}
    \caption{Projections at the start of the flat top (circle) and just before the time of disruption (cross), for a set of $\approx$200 discharges corresponding to the ITER Baseline (IBL), density limit (DL) experiments and negative triangularity (NT) configurations. At the start, the discharges are clustered together in a region with low disruption risk. At $t_D$, they are split onto different peaks in regions of higher $D_{\textit{risk}}$.}
    \label{fig:disruptionspread}%
\end{figure}

\begin{figure}[t]
\begin{center}\includegraphics[width=1\linewidth]{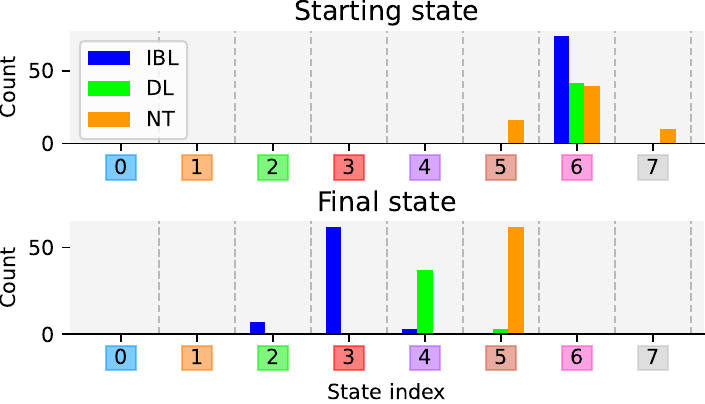}\end{center}
    \caption{The assignment of the projections from Figure~\ref{fig:disruptionspread} to the different states using $\mathcal{C}$ (Equation~\ref{eq:clustering}). While clustered together at the start of the flat top, there is a clear separation in the terminal states.}
    \label{fig:disruptionstate}%
\end{figure}

\begin{figure}[t]
\begin{center}\includegraphics[width=1\linewidth]{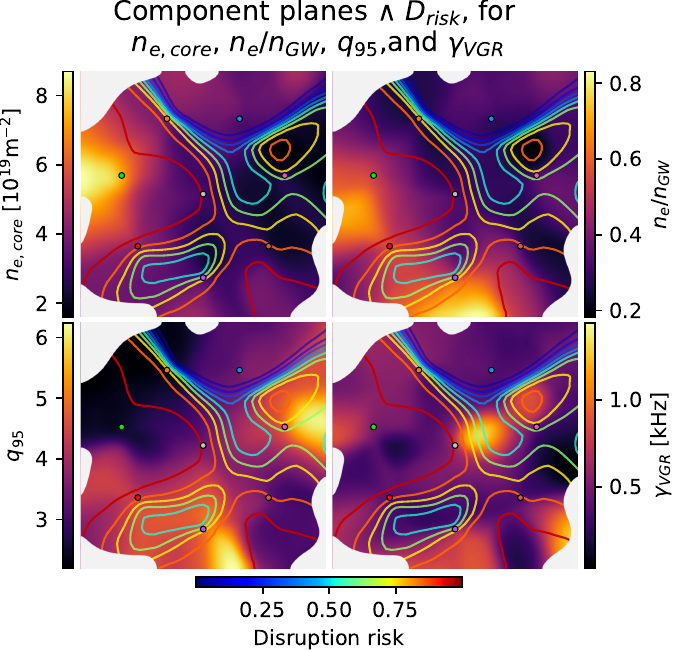}\end{center}
    \caption{Component planes (projected using $p_\theta(\mathbf{x}|\mathbf{z})$) with the disruption risk overlaid on top. By projecting back to data space we can interpret the correlations with disruptive regimes. For example, we see a low $q_{95}$ in regions of high $D_{\textit{risk}}$, also corresponding to the region of most IBL-related disruptions. Similarly, we find a high Greenwald fraction near the DL-related disruptions, and an elevated vertical growth rate near the NT-related disruptions.}
    \label{fig:componentplanes}%
\end{figure}
\textbf{Disruption clustering.} For evaluating the clustering of disruptions, we select a set of $\approx$200 disrupting shots that correspond to either the ITER Baseline (IBL) scenario ~\cite{labit2024}, density limit (DL) experiments~\cite{sieglin2025}, or negative triangularity (NT) configurations~\cite{coda2022}. For all these shots, we plot the projections at the start of the flat top (circle) and the last projection before the disruption\footnote{0 to \SI{5}{\milli\second} up to $t_D$, depending on the alignment of the sampling rate and $t_D$.} (cross) in Figure~\ref{fig:disruptionspread}. The initial states are clustered, whereas there is a clear distinction at the time of disruption, see also Figure~\ref{fig:disruptionstate} for the state assignment at the start and end of the latent trajectories. Additionally, we see that initial states are projected in regions with lower values for $D_{\textit{risk}}$, whereas there is a clear increase at the time of disruption. As such, $D_{\textit{risk}}$ seems to provide sensible correlations with known disruption proxies globally (Section~\ref{ss:explatentspace}), and locally different operational limits correspond to different regions.

\textbf{Component planes.} To better understand the clustering we can visualize the physics quantities as a function of $\mathbf{z}$ using decoder distribution $p_\theta(\mathbf{x}|\mathbf{z})$. These component planes, overlaid with the disruption risk, are depicted in Figure~\ref{fig:componentplanes}. We see correlations with the expected operational limit or characteristic quantities for the distinct categories of disruptions. For example, around the region of most IBL disruptions we find low values for $q_{95}$ ($\leq 4$), a peak in the Greenwald fraction for the DL disruptions ($\geq0.7$), and an elevated $\gamma_{\textit{VGR}}$ for the NT experiments, as a proxy for vertically unstable plasmas~\cite{marchioni2024,coda2022}.

\subsection{Disruption counterfactual 
analysis}\label{ss:expcounterfactual}
\begin{figure}[t]
\begin{center}\includegraphics[width=0.9\linewidth]{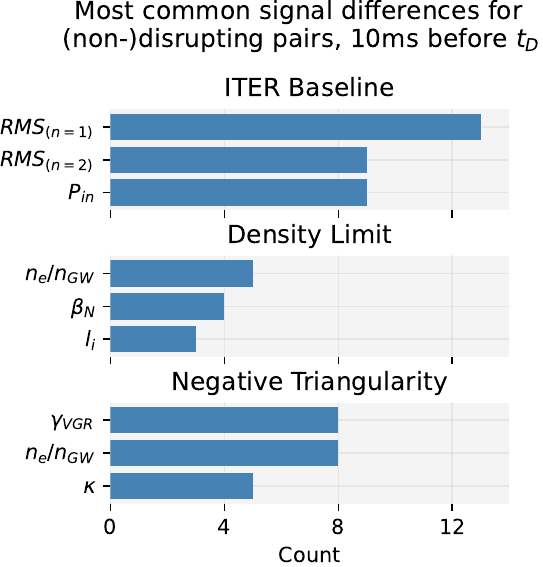}\end{center}
    \caption{Automatically identified parameters correlating to disruptions. Parameters are identified by finding differences in data distributions just before $t_D - \SI{10}{\milli\second}$ (Figure~\ref{fig:shot_pair_full}) for a disrupting discharge and the most similar timestep for a counterfactual non-disrupting match. The identified features generally fall in line with past studies, i.e., we find MHD-related instabilities in the IBL case~\cite{labit2024}, recover the Greenwald fraction as most important parameter for DL disruptions~\cite{sieglin2025} and find features associated with vertical instability for the NT scenarios~\cite{coda2022,marchioni2024}.}
    \label{fig:disrupionfeatures}%
\end{figure}

\begin{figure*}[t]
\begin{center}\includegraphics[width=0.88\linewidth]{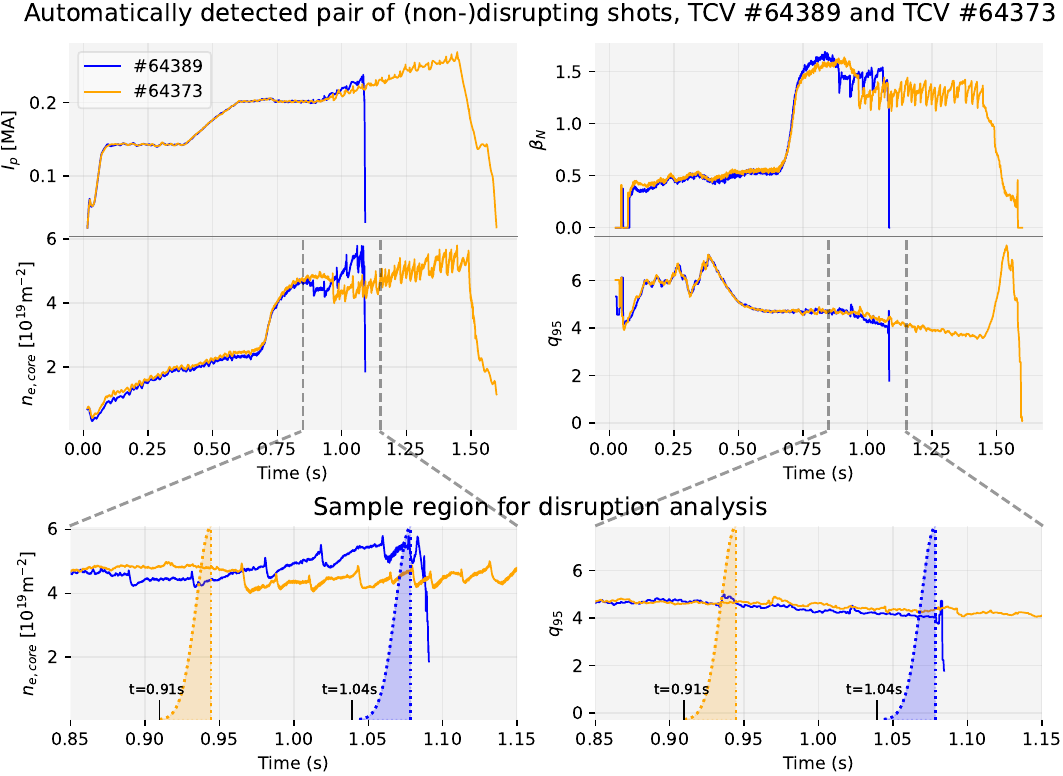}\end{center}
    \caption{An example discharge from IBL scenario development experiments that ended in a disruption, TCV \#64389, and its automatically identified counterfactual that did not disrupt, TCV \#64373 (top). We find the closest point in the latent space (Figure~\ref{fig:shot_pair_proj}) for \#64389 compared to \#64373 $\SI{10}{\milli\second}$ before $t_D$, and sample signal values in this window to identify significant differences between the two shots (bottom).}
    \label{fig:shot_pair_full}%
\end{figure*}

\begin{figure}[t]
\begin{center}\includegraphics[width=0.65\linewidth]{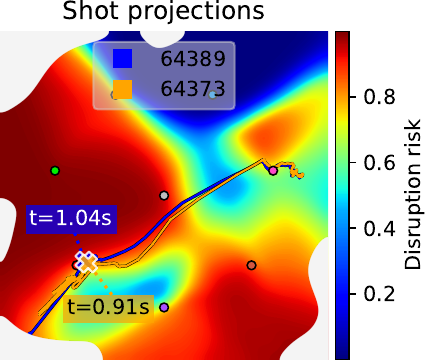}\end{center}
    \caption{Projections of TCV \#64389 and TCV \#64373 in the latent space (on top of $D_{\textit{risk}}$). We mark the position \SI{10}{\milli\second} before $t_D$ for \#64389 and the corresponding comparison point used for \#64373.}
    \label{fig:shot_pair_proj}%
\end{figure}

\begin{figure}[t]
\begin{center}\includegraphics[width=1\linewidth]{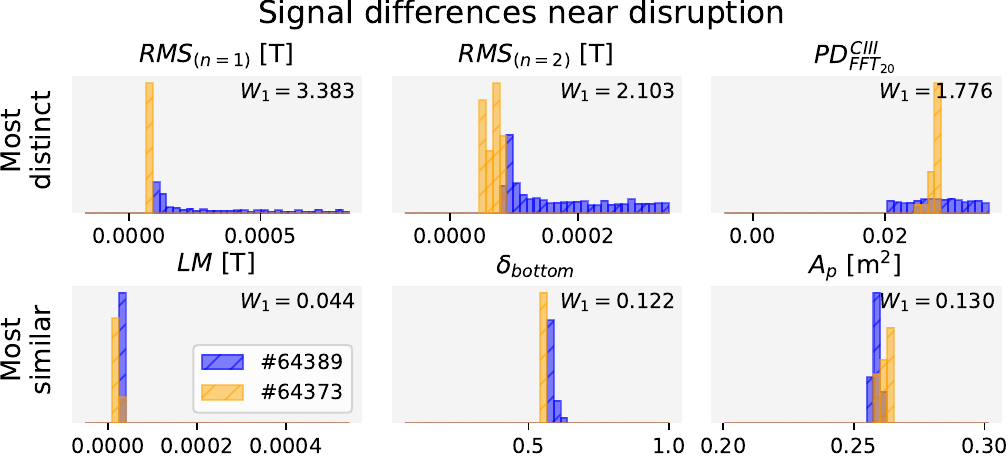}\end{center}
    \caption{Empirical distributions up to \SI{10}{\milli\second} before $t_D$ for the disrupting case and the identified closest point for the counterfactual case. We order the signals by 1-Wasserstein ($W_1$) distance on standardized feature values, and plot the top 3 most dissimilar and the top 3 most similar features.}
    \label{fig:shot_pair_hist}%
\end{figure}

\textbf{Counterfactual analysis.} To be useful for downstream tasks, the latent space should be suitable for identifying non-trivial connections between different discharges. To demonstrate this principle, we conduct a demonstrative study on identifying features connected to disruptions. Specifically, we use the model to identify pairs of discharges that are similar, but one ends in a disruption whereas the other ends in a regular termination. If the two are sufficiently similar otherwise, they can be used to do (approximate) counterfactual analysis~\cite{Lewis1986}. We additionally use the model to identify the most similar timestep in the non-disrupting shot, compared to \SI{10}{\milli\second} before $t_D$ for the disrupting case. Consequently, we can automatically identify parameters connected with disruptions in an interpretable manner.

\textbf{Procedure.} The low dimensionality of $\mathbf{z}$ allows us to efficiently compute distances between shots using Dynamic Time Warping (DTW)~\cite{dtw1978}. For each scenario and its disrupting discharges, we compute the optimal alignment w.r.t.\ all other discharges of the same category, and select the closest discharge as the counterfactual case. Then, we compute the closest point in the latent space between the disrupting shot $\SI{10}{\milli\second}$ before disruption and a timestep in the counterfactual shot, and sample parameters for both discharges using a half-Gaussian distribution with $\sigma=\SI{10}{\milli\second}$ at this point. See Figure~\ref{fig:shot_pair_full} for an example of an automatically identified pair and its corresponding sampling window, along with their projections and timestep-of-interest in Figure~\ref{fig:shot_pair_proj}.

By computing a distance metric on a normalized scale---standardizing feature values using the train set statistics---we can compare the relative differences in feature values. To quantify this difference, we utilize the Wasserstein distance~\cite{Kantorovich1960}, which can be interpreted as a cost of transforming one distribution to the other. We plot the most distinct and least distinct empirical feature distributions for the example case in Figure~\ref{fig:shot_pair_hist}. By placing a minimum threshold on the DTW distance between the shots and the Wasserstein difference between the signal distributions, we can robustly identify relevant features on a larger scale. Specifically, we select shots with a \textit{DTW} distance of at most 100 in latent-space scale (for reference, the visualized domain of $\mathbf{z}$ spans [-2.2, 2.2]), and a standardized Wasserstein distance $W_1 \geq 3$.

\textbf{Results.} We plot the most commonly identified significant feature difference for all IBL, DL and NT disruptions in Figure~\ref{fig:disrupionfeatures}. The identified features generally align with known limits. The IBL scenario performance is often limited by MHD instabilities on TCV~\cite{labit2024}, and consequently discrepancies in MHD activity are commonly the difference between (non-)disrupting shots. For the density limit, we expect the Greenwald fraction as the primary operational limit~\cite{sieglin2025}. Lastly, negative triangularity on TCV is known to be vertically unstable~\cite{coda2022,marchioni2024}, which is well captured by the vertical growth rate $\gamma_{\textit{VGR}}$ signal. 

Notably, by framing the automated analysis as identifying appropriate discharges and times therein, we can include additional signals not used in the model and compare different signals at arbitrary sampling rates, highlighting the flexibility of the approach.

\section{Conclusions and Discussion}\label{sec:conclusions}
We have presented a method that identifies a low-dimensional, abstract representation of the operational space of a tokamak. The method extends the VAE framework for properties desired for plasma state monitoring and disruption characterization. Specifically, the representation is optimized to model time series of discharge measurements as latent trajectories, to provide a clustering for distinct operating regimes, all while being informative w.r.t.\ the disruption risk. 

We have evaluated the method using a dataset of approximately 1600 TCV discharges modeling dynamics in the flat-top phase, covering flat-top disruptions and regularly-terminating shots. We investigated the identified latent variable w.r.t.\ the identified disruption-risk variable and its correspondence to the actual disruption rate and disruptivity computed after training. Additionally, we compared the identified states with known plasma state descriptions from external tools~\cite{poels2025ldh,defuseiaea}. We validated the ability to separate disruptions associated with distinct operational limits, as found in ITER Baseline scenarios, density limit experiments and negative triangularity configurations. Finally, we demonstrated the utility of the tool for downstream analyses by conducting an exploratory study on disruption-related features using counterfactual analysis.

\begin{figure}[t]
\begin{center}\includegraphics[width=0.65\linewidth]{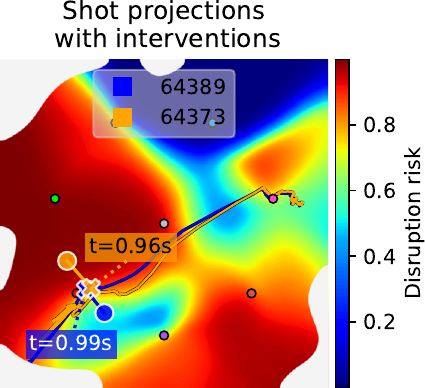}\end{center}
    \caption{Alteration of discharge projections just before the disruption towards a safer region (\#64389), or later in the discharge (\#64373) towards a less safe region.}
    \label{fig:alterationprojection}%
\end{figure}

\begin{figure}[t]
\begin{center}\includegraphics[width=1\linewidth]{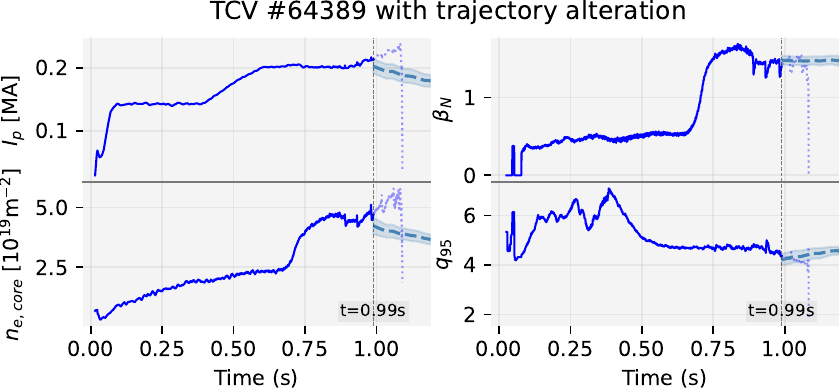}\\\vspace{0.2cm}\includegraphics[width=1\linewidth]{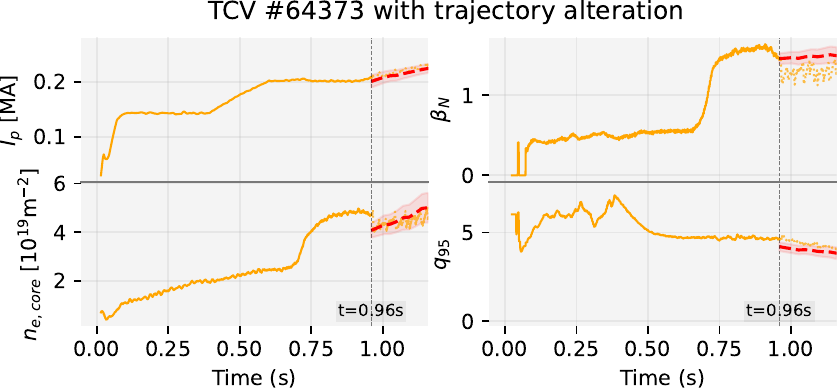}\end{center}
    \caption{Projections of the altered trajectories illustrated in Figure~\ref{fig:alterationprojection} in data space. Potentially, one could utilize a latent variable approach to project ongoing discharges in real time, to identify parameter changes gradually steering a discharge towards less disruption-prone regimes.}
    \label{fig:alterationreal}%
\end{figure}

The main downside of the approach is the relatively slow timescales that are represented in the latent space. The limited capacity of a low-dimensional latent variable coupled with the large variety in plasma scenarios means only a limited amount of information can be represented. However, a particular focus on the fast timescales around the time of disruption is of interest for analysis regarding the associated chain-of-events.

\subsection{Future work}
Modeling fast timescales requires an increased expressivity of the generative model. Likely, a trade-off has to be made between interpretability and expressivity, given that the 2D latent variable forms one of the main bottlenecks in this regard. Potential approaches include increasing the latent space dimensionality, extending the latent variable to a hierarchical structure~\cite{hvae2020} or utilizing alternative latent variable formulations, e.g. based on normalizing flows~\cite{horvat2021normalizing}.

To extend towards multi-machine analysis, it is of interest to learn a single latent representation found through discharges coming from different tokamaks. Such an approach comes with the challenge of properly integrating dynamics that operate on different time and spatial scales---a latent variable that simply clusters each device into a separate region provides little benefit over learning a representation on a per-device basis. Notably, prior works have studied multi-machine disruption prediction~\cite{katesharbeck2019,zhu021disr,zheng2023disr} and multi-machine representation learning in non-disruption contexts~\cite{aaromultimachine}, providing a basis for this extension.

Finally, the latent space-approach could be utilized in the setting of plasma control. For example, one could project a discharge onto $\mathbf{z}$ in real time as new measurements come in, and explore the surrounding region to inform control targets. One could project in the direction of a lower disruption risk to find physics quantities close to the current regime that are connected to less disruption risk. See Figure~\ref{fig:alterationprojection} as an example of such an intervention for \#64389, along with Figure~\ref{fig:alterationreal} for the corresponding quantities in data-space. Or to the contrary, one could project towards regions of higher disruptivity to inform a controller of parameter spaces to avoid, see \#64373 in Figures~\ref{fig:alterationprojection} and~\ref{fig:alterationreal}. Additionally, one could extend the formulation to include a forward model that predicts future states based on control actions~\cite{kitaug}, providing even more information for advanced control schemes~\cite{vu2021,galperti2024}.

\section*{Acknowledgements}
This work was funded in part by a Swiss Data Science Center project grant (C21-14). This work has been carried out within the framework of the EUROfusion Consortium, partially funded by the European Union via the Euratom Research and Training Programme (Grant Agreement No 101052200 — EUROfusion). The Swiss contribution to this work has been funded in part by the Swiss State Secretariat for Education, Research and Innovation (SERI). Views and opinions expressed are however those of the author(s) only and do not necessarily reflect those of the European Union, the European Commission or SERI. Neither the European Union nor the European Commission nor SERI can be held responsible for them. This work was supported in part by the Swiss National Science Foundation. This work used the Dutch national e-infrastructure with the support of the SURF Cooperative using grant no. EINF-7709.

\section*{References}
\bibliographystyle{plainurl_abrev}
\bibliography{main}

\clearpage
\appendix
\renewcommand{\thesection}{\appendixname~\Alph{section}}
\onecolumn
\setcounter{footnote}{1}
\counterwithin{figure}{section}
\counterwithin{lstlisting}{section}
\renewcommand{\thefigure}{\Alph{section}.\arabic{figure}}
\renewcommand{\thetable}{\Alph{section}.\arabic{table}}

\section{Settings and hyperparameters}\label{ap:vaeparameters}
\textbf{Training.} The model training procedure consists of minimizing the combined loss function defined in Equation~\ref{eq:totalloss}. An optimization step consists of sampling a sequence of input timewindows $\mathbf{x}^{t_{m-w}:t_m} \in \mathbb{R}^{U \times w}$ and disruption risk labels $y^{t_m} \in \mathbb{R}$. Specifically, we gather sequences of inputs:
\begin{align}
\left\{ \left( \mathbf{x}^{t_{m_i - w} : t_{m_i}},\, y^{t_{m_i}} \right) \right\}_{i=1}^{N}, \quad \text{where } m_i = m_0 + i s,
\end{align}
for arbitrary starting timestep $m_0$, from which we iterate with stride $s$ for sequences of length $N$. Then, we iteratively produce outputs following the procedure described in Section~\ref{ss:mmvae}. Notably, for the first timestep we do not have a previous latent position $\boldsymbol{\mu}$ available. Here, we use a separate encoder $\boldsymbol{\mu}^{t_m} = f_{\phi,0}(\mathbf{x}^{t_{m-w}:t_m})$ that is otherwise identical to $f_{\phi}$ (Equation~\ref{eq:encoderfullmodel}), with the exception of dropping the dependence on the previous state. We do not update model parameters using the entire sequence at once, but rather iteratively detach the gradients during this procedure. The main dataset parameters are repeated in Table~\ref{tab:data_params}, whereas the optimization-related settings are provided in Table~\ref{tab:training_params}. The weights of the loss terms are $a$ = 2, $b$ = 5, $c$ = 100 and $d$ = 10, respectively. Details on the choice of all hyperparameters is given below in \textit{hyperparameter choice}.

\textbf{Distribution parameters.} For the prior, we scan a set of structures for a single Gaussian centered at (0, 0), surrounded by a set of Gaussians spaced equidistant around this center point. We also explored optimizing the structure of the prior, but found that the procedure often led to undesirable solutions, such as several peaks merging into one or peaks completely separating from each other. Consequently, we leave the adaptive-prior setting for future work.

\begin{figure}[h]
\begin{center}\includegraphics[width=0.35\linewidth]{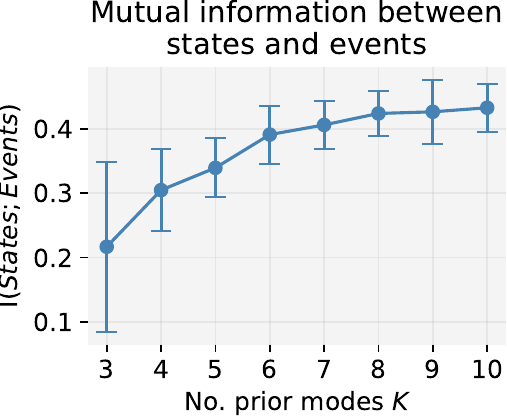}\end{center}
    \caption{The mutual information between event detections and the plasma state assignment $\mathcal{C}$, for a variety of structures for $p(\mathbf{z})$; we aggregate over other varied model parameters, indicated by the error bars (standard deviation). Each prior consists of one Gaussian centered at (0, 0), with $K-1$ Gaussians spaced equidistant around this center point.}
    \label{fig:mi_states}%
\end{figure}

\begin{figure}[h]
\begin{center}\includegraphics[width=0.65\linewidth]{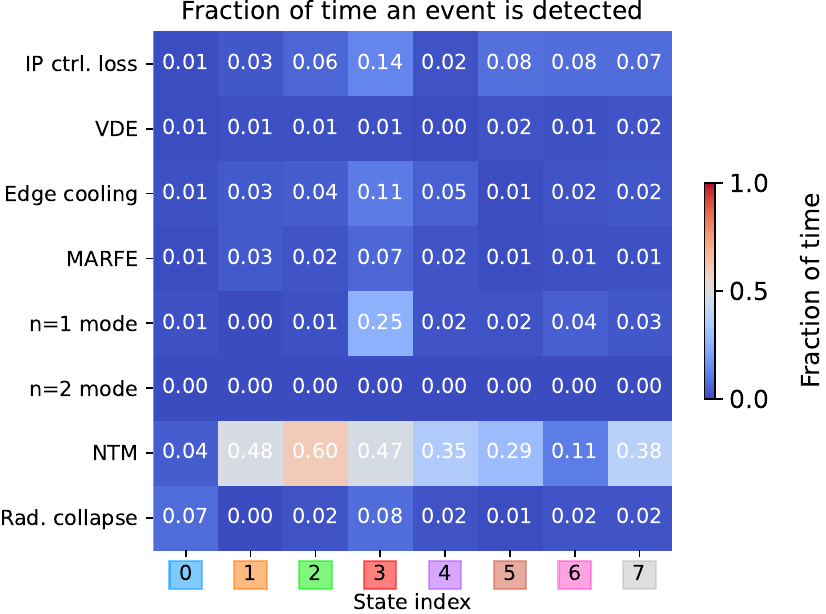}\end{center}
    \caption{Correlation between event detections computed using~\cite{defuseiaea} and the states found in $\mathbf{z}$. With the exception of NTMs, most states are only detected for small fractions of the total plasma duration. Most likely, the disruption-related events operate on smaller timescales than the global plasma regimes modeled by the latent variable $\mathbf{z}$.}
    \label{fig:heatmap_events}%
\end{figure}

To quantify the effect of adding prior modes, we compute the mutual information~\cite{mutualinformation} between the identified states and a set of event detections~\cite{defuseiaea,poels2025ldh} (Figures~\ref{fig:heatmapldh} and~\ref{fig:heatmap_events}). This metric can be interpreted as the extent to which knowledge of the state helps with estimating the odds of observing an event. We compute this metric for a range of $K \in [3, 10]$ prior modes and aggregate over all other varied parameters in our search, see Figure~\ref{fig:mi_states}. Here, we see that the added information starts to level off at around $K=8$ components, and subsequently select this quantity for our model.

Remaining distribution hyperparameters are provided in Table~\ref{tab:distribution_params}. Lastly, we provide details on the cluster assignment, where the implementation slightly deviates from the likelihood formulation. We use the inverse distance to the prior means with a scaling hyperparameter, which is equivalent for the most likely component, but reformulates the spread for the full distribution. The implementation is as follows:
\begin{align}\label{eq:clustering}
    \mathcal{C}:= p(\mathcal{C}=k|\mathbf{z}) 
= \frac{\exp\left(\frac{1}{\tau\,(1 + \|\mathbf{z}-\boldsymbol{\mu}_k\|^2)}\right)}
{\sum_{j=1}^{K}\exp\left(\frac{1}{\tau\,(1 + \|\mathbf{z}-\boldsymbol{\mu}_j\|^2)}\right)},
\end{align}
which is equivalent to the Softmax of inverse distances, with 1 added to the denominator for numerical stability as the distance goes to 0, and a logistic scaling parameter $\tau$ as hyperparameter. We set $\tau$ to 0.05 for all models.

\textbf{Hyperparameter choice.} Optimizing the model hyperparameters is challenging due to the many conflicting objectives. Additionally, it is nontrivial to capture the quality of the identified latent space in a single metric, which limits the ability to use automated parameter optimization tools. Instead, we systematically scan a variety of parameter ranges for the disruption risk parameters ($A \in \{0.05, 0.1, 0.15\}$\SI{}{\second}, $B \in \{0.1, 0.5, 1.0\}$\SI{}{\second}) and prior modes ($K \in [3, 10]$). Additionally, we scale the neural network architecture parameters by $\{\cdot \frac{1}{4}, \cdot \frac{1}{2}, 1\}$. Loss weights $a$, $b$, $c$ and $d$ are tuned by hand. Models were evaluated by manually inspecting the latent space distributions, and subsequently selecting a model by three properties: (1) smooth projections of sequential measurements; (2) assigning some probability density to different peaks; and (3) clear separation w.r.t.\ the disruption risk variable. Unfortunately, manual parameter optimization means we are likely selecting suboptimal settings, making the development of quantitative measures for `latent space quality' of interest for future works. 

An overview of the specific model architectures and parameters is given in Table~\ref{tab:encoder_static} for the initial-timestep encoder, Table~\ref{tab:encoder_rnn} for the regular encoder, Table~\ref{tab:decoder} for the decoder and Table~\ref{tab:drisk} for the disruption risk network.

\begin{table}[H]
\centering
\begin{tabular}{p{3.5cm}p{6cm}}
Parameter & Value \\
\cmidrule[\heavyrulewidth]{1-2}
Stride $s$ & 50 \\
Timewindow size $w$ & 50 \\
1-labels before $t_D$, $A$ & \SI{0.15}{\second} \\
0-1 ramp before $t_D$, $B$ & \SI{1.0}{\second} \\
\end{tabular}
\caption{Data-related hyperparameters.}
\label{tab:data_params}
\end{table}

\begin{table}[H]
\centering
\begin{tabular}{p{3.5cm}p{6cm}}
Parameter & Value \\
\cmidrule[\heavyrulewidth]{1-2}
Optimizer & Schedule-free Adam~\cite{defazio2024} \\
Optimizer learning rate & 0.005 \\
Optimizer warmup steps & 50 \\
Unroll steps ($N$) & 200 \\
Gradient detach interval & Every 25 steps \\
Epochs & 100 \\
Batch size & 512 \\
\end{tabular}
\caption{Training-related hyperparameters.}
\label{tab:training_params}
\end{table}

\begin{table}[H]
\centering
\begin{tabular}{p{3.5cm}p{6cm}}
Parameter & Value \\
\cmidrule[\heavyrulewidth]{1-2}
Radius of surrounding prior components& 1.5 \\
Total number of prior components & 8 \\
Prior variance $\sigma^2_{\text{p}}$ & 0.1 \\
Encoder variance $\sigma^2_{\phi}$ & 0.03 \\
Decoder variance $\sigma^2_{\theta}$ & 0.05 \\
\end{tabular}
\caption{Distribution-related hyperparameters.}
\label{tab:distribution_params}
\end{table}

\begin{table}[H]
\centering
\begin{tabular}{p{3.5cm}p{6.5cm}}
Layer & Details \\
\cmidrule[\heavyrulewidth]{1-2}
\textbf{Input} & $\mathbf{x}^{t_{m-w}:t_m} \in \mathbb{R}^{U \times w}$ \\
FNO Layer & 20 $\rightarrow$ 64 channels, 8 modes, ReLU activation \\
FNO Layer & 64 $\rightarrow$ 64 channels, 8 modes, ReLU activation \\
Max pooling & Kernel size of 2 \\
Fully connected & 1600 $\rightarrow$ 2 \\
\textbf{Output} & $\boldsymbol{\mu}^{t_m} \in \mathbb{R}^2$
\end{tabular}
\caption{Architecture of the encoder for the initial timestep,  $\boldsymbol{\mu}^{t_m} = f_{\phi,0}(\mathbf{x}^{t_{m-w}:t_m})$.}
\label{tab:encoder_static}
\end{table}

\begin{table}[H]
\centering
\begin{tabular}{p{3.5cm}p{6.5cm}}
Layer & Details \\
\cmidrule[\heavyrulewidth]{1-2}
\textbf{Input}$_1$ & $\mathbf{x}^{t_{m-w}:t_m} \in \mathbb{R}^{U \times w}$ \\
FNO Layer & 20 $\rightarrow$ 64 channels, 8 modes, ReLU activation \\
FNO Layer & 64 $\rightarrow$ 64 channels, 8 modes, ReLU activation \\
Max pooling & Kernel size of 2 \\
Fully connected & 1600 $\rightarrow$ 128 \\
\textbf{Input}$_2$ & Concatenate $\boldsymbol{\mu}^{t_{m-s}} \in \mathbb{R}^2$ \\
MLP$_{\textit{out}}$ & [128 + 2] $\rightarrow$ 128 $\rightarrow$ 2, GELU activation~\cite{hendrycks2016} \\
\textbf{Output} & $\Delta\boldsymbol{\mu}^{t_m} \in \mathbb{R}^2$
\end{tabular}
\caption{Architecture of the residual encoder $\boldsymbol{\mu}^{t_m} = \boldsymbol{\mu}^{t_{m-s}} + f_{\phi}(\boldsymbol{\mu}^{t_{m-s}}, \mathbf{x}^{t_{m-w}:t_m})$.}
\label{tab:encoder_rnn}
\end{table}

\begin{table}[H]
\centering
\begin{tabular}{p{3.5cm}p{6.5cm}}
Layer & Details \\
\cmidrule[\heavyrulewidth]{1-2}
\textbf{Input} & $\mathbf{z}^{t_m} \in \mathbb{R}^2$ \\
Positional Encoding & $\gamma(\mathbf{z}^{t_m}) \in \mathbb{R}^8$, $L=2$ frequencies \\
MLP & 8 $\rightarrow$ 4096 $\rightarrow$ 4096 $\rightarrow$ 20, ReLU activation \\
\textbf{Output} & $\widetilde{\mathbf{x}}^{t_m} \in \mathbb{R}^{U}$ \\
\end{tabular}
\caption{Architecture of the decoder $\widetilde{\mathbf{x}}^{t_m} = f_\theta(\mathbf{z}^{t_m})$.}
\label{tab:decoder}
\end{table}

\begin{table}[H]
\centering
\begin{tabular}{p{3.5cm}p{6.5cm}}
Layer & Details \\
\cmidrule[\heavyrulewidth]{1-2}
\textbf{Input} & $\mathbf{z}^{t_m} \in \mathbb{R}^2$ \\
Positional Encoding & $\gamma(\mathbf{z}^{t_m}) \in \mathbb{R}^8$, $L=2$ frequencies \\
MLP & 8 $\rightarrow$ 512 $\rightarrow$ 1, ReLU activation \\
Output activation & Sigmoid \\
\textbf{Output} & $D_{\textit{risk}} \in \mathbb{R}$ \\
\end{tabular}
\caption{Architecture of the disruption risk network $D_{\textit{risk}} = f_D(\mathbf{z}^{t_m})$.}
\label{tab:drisk}
\end{table}

\newpage

\section{Extra results}\label{ap:vaeextraresults}

\end{document}